\documentclass[11pt]{article}
\usepackage[final]{acl}

\usepackage{times}
\usepackage{latexsym}
\usepackage[T1]{fontenc}
\usepackage[utf8]{inputenc}

\usepackage{microtype}

\usepackage{inconsolata}

\usepackage{graphicx}
\usepackage{makecell} % subfloats
\usepackage{type1cm} % type1 computer modern font
\usepackage{graphicx} % advanced figures
\usepackage{xspace} % fix space in macros
\usepackage{balance} % to better equalize the last page
\usepackage{booktabs} % nicer tables
\usepackage{multirow} % multi rows for tables
\usepackage{subcaption} % subfloats
\usepackage{bold-extra} % bold + {small capital, italic}
\usepackage[vlined,linesnumbered,ruled,noend]{algorithm2e} % algorithms
\usepackage{microtype} % compress text
\usepackage{siunitx} % \num for decimal grouping
\usepackage{xfrac} % nicer slanted fractions
\usepackage{mathtools} % amsmath++
\usepackage{url}
\usepackage{natbib}
\usepackage{rotating}
\usepackage{adjustbox}

\PassOptionsToPackage{hyphens}{url} % acmart compatibility
\PassOptionsToPackage{bookmarks, pdftex, colorlinks=true, pagebackref=true, backref=page}{hyperref} % acmart compatibility
\PassOptionsToPackage{square,numbers}{natbib} % acmart compatibility
\usepackage{cleveref} % smart references
\usepackage[hyperpageref]{backref} % back references
\usepackage{hyphenat} % name in single line
\usepackage{ragged2e}
\usepackage{CJKutf8} % Japanese

\usepackage[most]{tcolorbox}

\tcbset{
  promptbox/.style={
    colback=gray!3,
    colframe=black!60,
    fonttitle=\bfseries,
    coltitle=black,
    boxrule=0.5pt,
    arc=2mm,
    left=2mm,
    right=2mm,
    top=1mm,
    bottom=1mm,
    enhanced
  }
}

\usepackage[show]{chato-notes} % notes

\newenvironment{squishlist}
{\begin{list}{$\bullet$}
 {\setlength{\itemsep}{0pt}
     \setlength{\parsep}{3pt}
     \setlength{\topsep}{3pt}
     \setlength{\partopsep}{0pt}
     \setlength{\leftmargin}{1.5em}
     \setlength{\labelwidth}{1em}
     \setlength{\labelsep}{0.5em} } }
{\end{list}}

\renewcommand*\backref[1]{\ifx#1\relax \else (Cited on #1) \fi}

\graphicspath{{images}}

\title{Among Us: Language of Conspiracy Theorists on Mainstream Reddit}

\author{
 \textbf{Francesco Corso}\textsuperscript{1,2,3},
 \textbf{Giuseppe Russo\textsuperscript{4}},
 \textbf{Francesco Pierri\textsuperscript{1,*}},
 \textbf{Gianmarco De Francisci Morales\textsuperscript{2,3*}}
\\
\\
 \textsuperscript{1}Dipartimento di Elettronica, Informazione e Bioingegneria, Politecnico di Milano, Milan, Italy
 \\
 \textsuperscript{2}CENTAI, Turin, Italy
 \\
 \textsuperscript{3}Intesa Sanpaolo Innovation Center, Turin, Italy
 \\
 \textsuperscript{4}EPFL, Lausanne, Switzerland
 \\
 \textsuperscript{*}\small Equal contribution
 \small{
   \textbf{Correspondence:} \href{mailto:francesco.pierri@polimi.it}{francesco.pierri@polimi.it}
 }
}

\begin{document}
\newcommand{\R}[1]{\texttt{r/#1}\xspace}
\newcommand{\rconsp}{\R{conspiracy}}
\graphicspath{{images/}}

\maketitle
\begin{abstract}
The interaction between fringe subcultures and mainstream online communities poses significant challenges for understanding discourse on social media.
In this work, we investigate whether users active in conspiracy-focused communities exhibit detectable linguistic signatures when participating in general-interest spaces, such as news, humor, or hobbyist forums.
We analyze a large-scale longitudinal dataset of over 500 million comments spanning 10 years of Reddit activity, examining the communication patterns of these users across diverse social contexts independent of the topics they discuss.
We show that these users exhibit distinctive linguistic patterns that enable machine learning models to reliably distinguish them from the general population within individual communities (averaging 87\% accuracy across more than 20 binary classification tasks).
Crucially, no single aggregate model captures these patterns across communities, as community-specific models outperform global classifiers by up to 17 percentage points.
This result suggests that while these users are distinct, their linguistic expression is dynamic and highly responsive to the social norms of the environment they inhabit. 
Our findings suggest the need for tailored interventions in online spaces, as linguistic signals associated with conspiracy and fringe subcultures vary across communities and cannot be effectively addressed by uniform detection or moderation strategies.
\end{abstract}

\section{Introduction}
\label{sec:intro}

Social media platforms facilitate the rapid spread of information and the formation of communities around shared ideologies~\cite{loru2025ideology}. Within this ecosystem, mainstream narratives compete with alternative interpretations of reality~\cite{benkler2018network}, allowing subcultures to develop distinct epistemic norms. Among these, conspiracy theories have shown significant resilience and reach in online environments~\cite{starbird2017examining, monti_online_2023}.
Conspiracy theories are not merely fringe beliefs; they are influential, alternative narratives that explain events through the actions of secretive, malevolent groups~\cite{douglas_understanding_2019}. 
While often dismissed as speculation, their real-world consequences can be substantial: they have been linked to vaccine hesitancy and public health risks~\cite{enders2022relationship}, prompting official responses from public institutions.\footnote{\href{https://commission.europa.eu/strategy-and-policy/coronavirus-response/fighting-disinformation/identifying-conspiracy-theories_en}{EU Commission, 2021}} 
In more extreme cases, they can pose a threat to democratic institutions themselves---as exemplified by the January 6th, 2021 attack on the U.S. Capitol, which was partly fueled by conspiracy-driven rhetoric.\footnote{\href{http://www.washingtonpost.com/technology/2021/10/22/jan-6-capitol-riot-facebook}{Washington Post, 2021}}
These narratives can also serve as gateways to more radical ideologies, gradually reshaping individuals' perception of reality~\cite{basit2021conspiracy}. 

Despite extensive research on conspiracy theories on online platforms~\cite{tangherlini_automated_2020, faddoul_longitudinal_2020,samory2018spy,korenvcic2024overview,corso2025conspiracy,attanasio2025effects,corso2025androids}, how conspiracy theory believers express themselves in mainstream online spaces remains underexplored. 
Research suggests conspiracy theorists do not just hold different beliefs, but also communicate differently by employing specific rhetorical styles and vocabulary~\cite{samory2018spy}.
However, it remains unclear whether these linguistic patterns are confined to online conspiracy spaces or are intrinsic features of users' communication style, visible in mainstream discourse.

In this study, we investigate whether users active in conspiracy communities exhibit specific patterns of self-expression that distinguish them from ordinary users.
We use Reddit as our primary case study due to its large-scale, publicly accessible dataset and a unique structure based on topical communities, which enables the analysis of users' behavior across both fringe and conventional communities. 
We articulate our contributions within a comprehensive research question:
\textbf{Do users who engage with conspiracy theories display distinctive linguistic patterns in mainstream online communities, compared to users without such engagement?}

We employ a dataset of over 500 million comments and hundreds of thousands of users spanning over 10 years of activity on over 20 mainstream communities.
We then extract the psycholinguistic features of each comment, which are then aggregated into a user linguistic vector. 
We then use these vectors to train a supervised classifier, which shows remarkable accuracy and stability in distinguishing users active in conspiracy-related communities from those who are not.

By characterizing how users active in \rconsp adapt their language across communities, this study contributes to a more nuanced understanding of online conspiracy discourse, with implications for designing context-aware and proportionate interventions on digital platforms.

\section{Related Work}
\label{sec:related_w}
\subsection{Detection of Online Conspiracy Theories}
Computational approaches to detect online conspiracy theories have primarily focused on identifying explicit narrative structures and content signatures.
For instance, \citet{tangherlini_automated_2020} used structural modeling to map the narrative frameworks of theories such as ``Pizzagate'', distinguishing them from actual conspiracies based on the stability and interconnectivity of their subject-threat relationships. 
Similarly, \citet{samory_conspiracies_2018} employed topic modeling and n-gram analysis to capture the distinct ``conspiratorial language'' of dedicated forums, flagging content based on specific vocabulary and topic distributions rather than user style.
Early work by \citet{faddoul_longitudinal_2020} investigated the detection of conspiracy-related content in multimodal settings by leveraging textual information associated with videos, such as captions, snippets, and top comments. 
More recently, \citet{corso2025conspiracy} demonstrated the effectiveness of Large Language Models for identifying conspiratorial content on TikTok, highlighting the potential of recent advances in generative AI to support moderation efforts in multimodal platforms. 
Relatedly, \citet{diab2024classifying} explored the capabilities of LLMs for detecting conspiracy-related content on Reddit.

\subsection{User Pathways To Fringe Communities}
Beyond the analysis of content, recent work has modeled online participation in conspiracy discourse as a dynamic, gradual trajectory of engagement.
\citet{klein2019pathways} identified linguistic precursors of conspiracy activity in fringe communities, showing that users who eventually join conspiracy forums display distinct participation and linguistic markers in their posting history before their first direct participation.
\citet{phadke2022pathways} characterize the evolution of conspiracy engagement as a multi-stage process, wherein users progressively adopt the norms and vocabulary of extremist communities through sustained interaction. 
This effect is further explored by research on gateway communities; for instance, \citet{rollo2022communities} and \citet{habib2022} demonstrate how adjacent ideological spaces, such as the ``manosphere'', facilitate a user's drift toward more radical, anti-establishment narratives.

\subsection{The Language of Conspiracy Theories}
Psychological research consistently characterizes conspiracy theories as narratives that attribute hidden, intentional agency to powerful actors and that fulfill existential and social motives, particularly under conditions of uncertainty and threat~\cite{douglas_understanding_2019,douglas_what_2023}. 
Across domains, conspiracy theories are communicated through recurring linguistic patterns, including causal chaining, certainty and authority markers, attribution of malicious intent, and framing that contrasts an informed ingroup with a deceptive outgroup~\cite{van_prooijen_conspiracy_2017,meuer_how_2023}.
These communicative features are not incidental: field studies show that conspiratorial discourse is systematically more emotional and distrust-laden than non-conspiratorial explanations of the same events~\cite{fong2021language}.

\section{Methods}
\label{sec:Methods}
\subsection{Data Collection}
Our primary data source is the Pushshift Reddit dataset ~\cite{baumgartner2020pushshift}, a public archive of posts and comments on the platform, which excludes content created by deleted users or removed by moderators or the authors themselves.
Here, we focus on user comments as the fundamental unit of analysis, since they represent the most common form of interaction on Reddit and typically indicate active participation within a community~\cite{Naab2023}. 

We extracted the full archive of available comments from the largest conspiracy-focused subreddit, \rconsp, spanning from its creation on 25 January 2008 to 31 December 2023.
The resulting corpus includes approximately \num{25.7} million comments from over \num{980000} unique users, with an average of 5k daily comments.

\begin{figure}[t]
    \centering
    \includegraphics[width=0.9\linewidth]{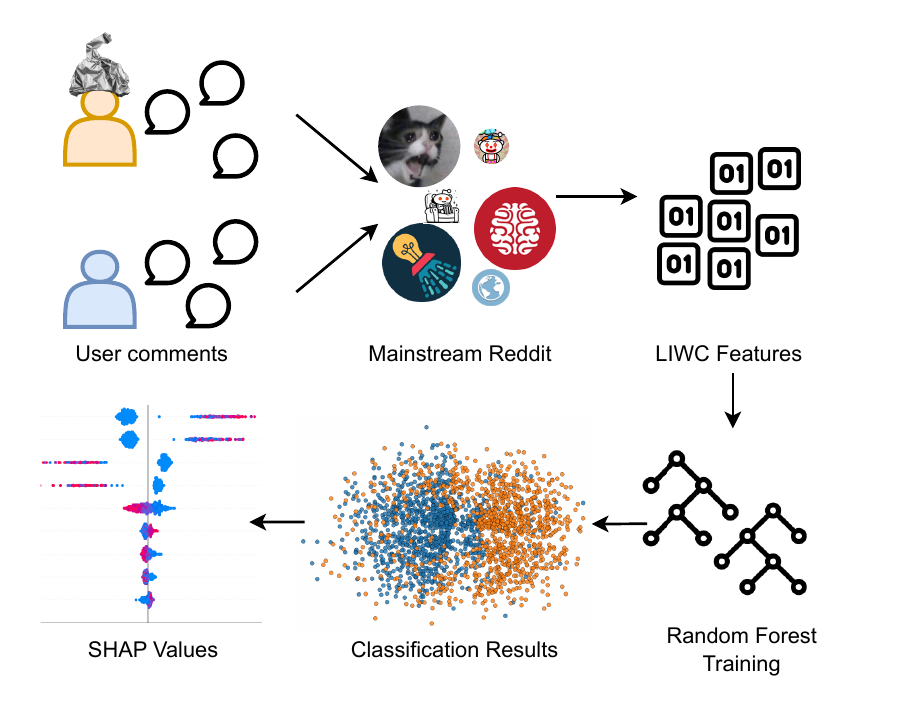}
    \caption{Diagram showing the workflow, from data collection to classification and feature analysis.}
    \label{fig:pipeline}
\end{figure}

We additionally collected the complete activity of the top subreddits ranked by the number of subscribed users as of June 2025.\footnote{\url{https://www.reddit.com/best/communities/1}}
We selected 22 of the top communities, excluding the following three subreddits: \texttt{r/AskReddit} and \texttt{r/worldnews}, due to their broad and heterogeneous content, and \texttt{r/gaming}.
This selection strategy allowed us to observe the behavior of conspiracy users outside of \rconsp while reducing the bias introduced by the uneven activity distribution of conspiracy users across mainstream subreddits.
We found instead that selecting subreddits based purely on where \rconsp users were most active outside their primary community (a case of selection on the outcome) produced an anomalous, non-decreasing activity distribution---unlike the typical long-tailed activity distributions seen on Reddit and other platforms~\cite{valensise2019drifts}.
Additional details on this selection rationale are provided in \Cref{A1}.

We limited our analysis to the period from 2013 to 2023 (inclusive), discarding earlier activity.
We made this decision because several mainstream subreddits considered, such as \texttt{r/ShowerThoughts} (created on February 5, 2012), were established well after the creation of \rconsp.

The final dataset contains approximately \num{510} million comments.
From this corpus, we removed comments authored by known bots or suspicious accounts using predefined lists~\cite{rollo2022communities}.
Furthermore, for each mainstream subreddit, we excluded users who posted fewer than \num{20} comments within that community.
This threshold was applied to ensure that our analysis includes users with a strong and consistent signal of engagement.
A more fine-grained breakdown of the collected data is presented in \Cref{tab:sizes_comments} in the Appendix.

\subsection{Psycholinguistic Features and Dataset Construction}

To characterize the psycholinguistic patterns in user comments from the selected mainstream subreddits, we computed user-level embeddings based on LIWC-22~\cite{boyd2022development}, a widely employed dictionary-based tool that extracts linguistic features associated with psychological, cognitive, and social processes. 
LIWC has been extensively applied in psychological and social media research to infer user traits and mental states from language use~\cite{silva2021predicting, tausczik2010psychological}.
We applied the standard LIWC-22 processing pipeline to each comment collected during the data acquisition phase, resulting in \num{115} linguistic and psycholinguistic features per comment. 
We excluded five features related to punctuation, retaining \num{110} features for subsequent analysis.

To derive a single representation for each user within each mainstream subreddit, we computed the mean feature vector across all comments posted by that user in that community.
This aggregation yields a user-level embedding that captures their overall psycholinguistic profile based on language use within the subreddit.
This method allowed us to create comprehensive user representations that capture psycholinguistic traits and provide a unified framework for comparing language patterns across conspiracy and non-conspiracy users.

\subsection{Experimental Design}

We operationalize our research question by evaluating the ability of a Random Forest model to distinguish between the linguistic features of users active in the \rconsp subreddit and those of users who have not engaged with \rconsp, based on their language use within \emph{mainstream subreddits}.
Given the structured and interpretable nature of our feature space, Random Forests provide a robust modeling approach~\cite{haddouchi2024survey}.
Moreover, our objective in this study is not to build a high-performance classifier per se, but to use the classifier as a proxy to measure the distinguishability of language.

The rationale behind this procedure is grounded in prior literature: several studies suggest that conspiracy theorists are influenced by a ``conspiratorial mindset'', a psychological predisposition to interpret the events happening in the world as the result of hidden machinations orchestrated by secret, malicious entities~\cite{sutton_conspiracy_2020}.
The presence of this mindset is tied with the concept of the ``Monological World View''~\cite{sutton201414, swami2011conspiracist}, which suggests that believers in conspiracy theories perceive events in their lives through a markedly different lens compared to other individuals.
This worldview, a defining feature of conspiracy theorists, is thought to pervade all aspects of their lives~\cite{miani_interconnectedness_2022}, leading us to hypothesize that it can also manifest in their linguistic patterns. 
Though we can not prove that the online users under study possess this type of mindset, our work offers meaningful support to this hypothesis.

In this work, we use the term conspiracy-engaged users to refer specifically to users who have interacted with the subreddit \rconsp.
Thus, this operationalization captures observable engagement rather than underlying beliefs, and may overrepresent users willing to affiliate with this label while underrepresenting conspiratorial discourse expressed elsewhere.
Users are assigned to the positive class if they authored at least one comment in the \rconsp subreddit; all other users are assigned to the negative class. This inclusive operationalization is intended to capture a broad set of users who engage with conspiracy-focused content; as shown in later analyses, restricting the positive class to higher levels of engagement yields comparable results.

For each mainstream subreddit, we represent each user with a single psycholinguistic feature vector computed by averaging LIWC-22 features across all of their comments within that subreddit.
Thus, each user-subreddit pair constitutes one instance in the corresponding classification task.
To construct the dataset for each mainstream subreddit, we first collect all users in the positive class who meet the activity threshold for that subreddit. 
We then randomly sample an equal number of negative-class users from the same subreddit to obtain a balanced dataset. 
To account for variability introduced by negative-class sampling, we repeat this process five times, resulting in five replicas that share the same positive-class users but differ in their negative-class composition.
For each replica, we split users into disjoint training (80\%) and test (20\%) sets, ensuring that no user appears in both splits. 
Feature normalization and hyperparameter tuning are performed exclusively on the training data, using grid search with five-fold cross-validation. 
The final model is then evaluated on the held-out test set.
We apply this procedure independently for each of the five replicas and for each mainstream subreddit, yielding multiple evaluations per subreddit that reflect uncertainty due to negative-class sampling.

% Due to the balanced nature of the datasets we employed, accuracy serves as a reliable and meaningful evaluation metric for this task. 
Since both classes are represented in equal proportions, accuracy provides a clear measure of the classifier’s effectiveness without being skewed by class imbalances, making it a suitable choice for assessing how well the classifier distinguishes between conspiracy and non-conspiracy users. 
Additionally, using accuracy allows for straightforward comparisons across different models and experimental setups, reinforcing the robustness of our findings.

For each classification task, we assess statistical significance by using a permutation test~\cite{ojala2010permutation} in which training labels are randomly shuffled, and the model is retrained to generate a null distribution of performance (accuracy) scores. 
The performance of the original model is then compared against this distribution to test the null hypothesis of independence between features and labels.
We perform \num{100} label permutations on the training data.
The p-value is computed as $\frac{C + 1}{n_{perm} + 1}$, where $C$ is the number of permuted models that outperform the original model and $n_{perm}$ is the number of permutations. 
With \num{100} permutations, the minimum attainable p-value is \num{0.0099}.

% Model robustness was assessed using permutation testing~\cite{ojala2010permutation}, which yielded statistically significant results ($p < 0.01$) across all subreddits, indicating that model performance reflects meaningful patterns rather than random chance.

\subsection{Feature Importance}
In the context of Random Forest models, feature importance quantifies the contribution of each input variable to the predictive performance of the model. 
This metric is crucial for understanding which features are most influential in driving the model's decisions.
Traditionally, feature importance in Random Forests is assessed via metrics such as Gini importance or mean decrease in impurity.
However, these methods can sometimes be biased or difficult to interpret in the presence of correlated features.
To address these limitations, SHAP (SHapley Additive exPlanations) values offer a more robust and interpretable approach, providing a unified measure of feature importance by considering the contribution of each feature to the prediction of individual instances or a group of instances~\cite{lundberg2017unified}.
We first gather all the models produced in the main study and compute the absolute mean SHAP values for each model. 
This process results in a \num{110}-dimensional feature vector for each model, corresponding to the same psycholinguistic features employed to represent users in the classification experiment.
Each value in these vectors reflects the global importance of a given feature, calculated as the mean absolute SHAP value for that feature across all samples.
To obtain these SHAP values, we sample \num{700} positive instances from the dataset. 
As a result, for each subreddit, we derive a vector representing the importance of each classification feature within that subreddit, effectively mapping it into a 110-dimensional feature space.
Finally, we measure similarity between these vectors using cosine similarity, a robust metric in this context. 
Given the construction of these vectors, cosine similarity is particularly well-suited, as it prevents issues arising from scalar multiples.
The clustering in \Cref{fig:s1_features} is computed with the UPGMA hierarchical clustering algorithm~\cite{sokal1958statistical}.

\subsection{Additional Subreddits as Positive Class}
To assess the robustness of our findings, we extend our analysis to two additional communities on Reddit: \texttt{r/AskReddit} and \texttt{r/MensRights}.
\texttt{r/AskReddit} is one of the platform’s largest and most active subreddits, where users post open-ended questions intended to spark broad discussions.
Its generalist nature attracts a diverse user base, making it a useful benchmark for typical engagement patterns on Reddit.
\texttt{r/MensRights}, by contrast, is a more ideologically focused community that discusses issues perceived to affect men, including legal bias, custody disputes, and gender norms~\cite{decandia2022social}.
With over \num{300000} subscribers, it is also part of the broader ``manosphere'', a network of online communities that has been shown to overlap with conspiracy discourse and anti-establishment narratives~\cite{mamie2021anti}.
Similarly to \rconsp, \texttt{r/MensRights} is a longstanding community that has sparked controversy for its content but has not yet been taken down by Reddit moderators. 
This is the main reason we chose this community for the comparison against \rconsp.
In two separate experiments, we apply the same classification pipeline used for \rconsp. 
We analyze the language of users active in \texttt{r/AskReddit} and \texttt{r/MensRights}, based on their participation in top mainstream subreddits. 
Our goal is to determine whether the linguistic patterns of these users mirror those observed among conspiracy-engaged individuals---specifically, whether a machine learning model can reliably distinguish them from control users.

\subsection{Controlling for socio-demographic effects}
In designing our experiments, we exclude the potential influence of users’ socio-demographic attributes on the results.
We aim to ensure that the observed results can be attributed to conspiratorial engagement rather than background characteristics of the individuals. 
To test the plausibility of this assumption, we construct socio-demographic embeddings for all users active in \rconsp.
This is achieved by extracting the complete set of comments produced by these users on Reddit, totaling more than two billion posts.
For each user, we then rely on the subreddit embedding model introduced by \citet{waller2021quantifying}.
These embeddings act as proxies for latent socio-demographic attributes (e.g., age, gender, political orientation), inferred from patterns of community co-participation rather than directly observed traits.
We then use these embeddings to measure the difference in socio-demographic distribution across user pools. 
Our results show no meaningful correlation between these differences and changes in the accuracy of classification, thus providing empirical support for the decision to exclude socio-demographic attributes from the design of our experiments.
More details on these analyses can be found in \Cref{app:sociodemo}.
\section{Results}
\label{sec:results}

\begin{figure*}[!t]
    \centering
    \includegraphics[width=1\linewidth]{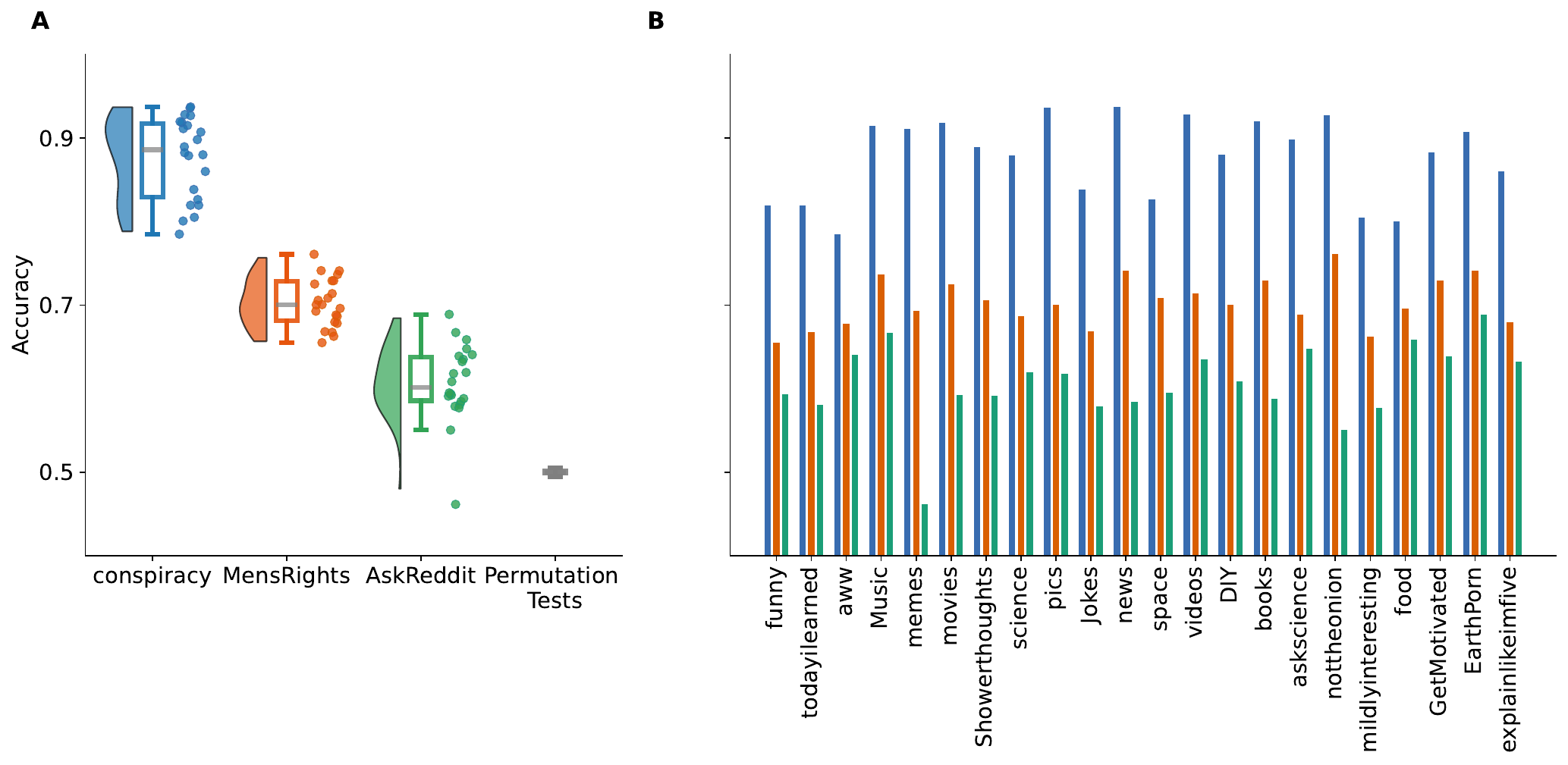}
    \caption{Results of the classification experiments across mainstream subreddits.
\textbf{A:} Distribution of classification accuracies in distinguishing users who engage with conspiracy communities (in blue) and two control communities (\texttt{MensRights} in orange, \texttt{AskReddit} in green) achieved by a Random Forest classifier, trained and evaluated separately on each mainstream subreddit; each dot represents a subreddit (Median = \num{0.87}, Min = \num{0.78}, Max = \num{0.95} for \texttt{conspiracy}; Median = \num{0.69}, Min = \num{0.65}, Max = \num{0.76} for \texttt{MensRights}; Median = \num{0.60}, Min = \num{0.46}, Max = \num{0.68} for \texttt{AskReddit}.
). Accuracy from randomized permutation tests is also shown in grey (Median accuracy = \num{0.5}).
\textbf{B:} Classification accuracy of the Random Forest model broken down by subreddit, ordered by popularity (number of subscribers) in descending order, for the conspiracy community (in blue) and the two control communities (MensRights in orange, AskReddit in green).}
    \label{fig:s1_both}
\end{figure*}

\subsection{Binary Classification Experiments}
\Cref{fig:s1_both} presents the general results of the binary classification experiments across all subreddits.
The median accuracy of a Random Forest classifier across subreddits is \num{0.87} (min = \num{0.78}, max = \num{0.95}), \num{37} p.p. higher than a random baseline classifier.
This performance demonstrates that the classifier can effectively differentiate between the two user groups, highlighting that individuals active in conspiracy communities exhibit distinct psycholinguistic patterns compared to users in mainstream subreddits.
These results hold for users who posted at least one comment in \rconsp.
Varying the level of activity within the conspiracy community---from 10 to more than 100 comments---yields comparable classification performance (see \Cref{app:disagg}).
%Taken together, these findings suggest that users who engage with a conspiracy community exhibit distinctive linguistic patterns relative to mainstream users from the time of their initial participation.

We test the robustness of the main results with three additional sets of experiments.
First, we perform a randomized permutation test to determine whether the classification accuracy could be attributed to chance.
As shown in panel A of \Cref{fig:s1_both}, the permutation test yields a median accuracy close to that of a random baseline classifier (\num{0.5}).
The results are statistically significant for every subreddit ($p < 0.001$), confirming that the model's performance reflects meaningful patterns in the data rather than random variation.

Next, we extend our analysis by constructing alternative positive classes using two additional target communities: \texttt{r/MensRights} and \texttt{r/AskReddit}.
This robustness check tests whether the identified psycholinguistic differences are unique to conspiracy communities or whether they also emerge in other ideologically-adjacent or general-interest forums.
As shown in panel A of \Cref{fig:s1_both}, the classifier performs significantly worse in these settings compared to the original experiment using \rconsp as the positive class (Mann-Whitney test, $p < 0.001$), thus indicating that the linguistic signals associated with conspiracy engagement are more distinctive than those emerging from general-interest or ideologically-adjacent communities.
Interestingly, the performance of the classifier is higher on \texttt{r/MensRights} than on \texttt{r/AskReddit}.
The reason for this difference might be due to the similar extremist nature of \texttt{r/MensRights}: both conspiracy and manosphere users believe they are privy to hidden truths (e.g., the ``red pill'' metaphor)~\cite{van2021digesting}.
These results reinforce our central claim that users who engage with conspiracy content exhibit language patterns that are significantly different from those of mainstream users.
%Additional details about this robustness check can be found in \Cref{sec:Methods}.

\begin{figure*}[t]
    \centering
    \includegraphics[width=1\linewidth]{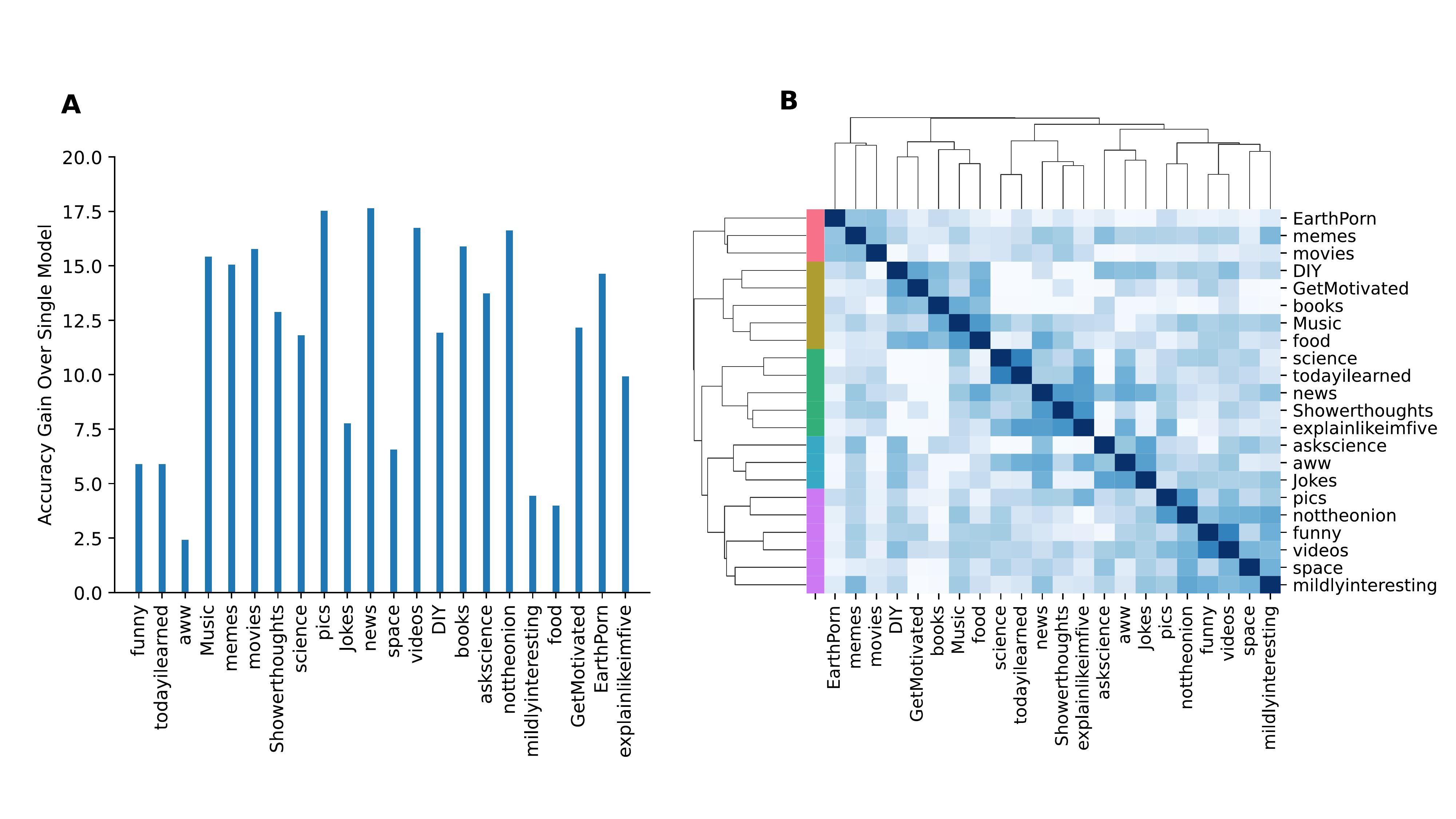}
    \caption{Analysis of linguistic fingerprints across different subreddits. \textbf{A:} Difference in accuracy (percentage points) between models trained separately for each subreddit and a single aggregate model trained on all subreddits. \textbf{B:} Clustering of mainstream subreddits based on the most important linguistic features used in the classification task. }
    \label{fig:s1_features}
\end{figure*}

Finally, we compare the performance of separate classifiers to a model trained on the combined data from all subreddits. 
While this aggregate model is more accurate than a random classifier baseline (accuracy = 0.76), we observe a drop in performance compared to subreddit-specific models.
As shown in \Cref{fig:s1_features}, panel \textbf{A}, the difference in accuracy with respect to the individual models ranges from \num{2.5} to \num{17} percentage points.
This result suggests that conspiracy-engaged users adapt their language to the specific communities they participate in, rather than maintaining a uniform conspiratorial discourse across contexts. 
This finding aligns with established research showing that online users adjust to the linguistic norms of individual communities~\cite{danescu2013no, zhong2017wearing}, reinforcing the idea that conspiratorial tendencies are shaped by the social and discursive environments of each subreddit.

%\label{app:feature_importance}
\begin{figure}[t]
    \centering
    \includegraphics[width=1\linewidth]{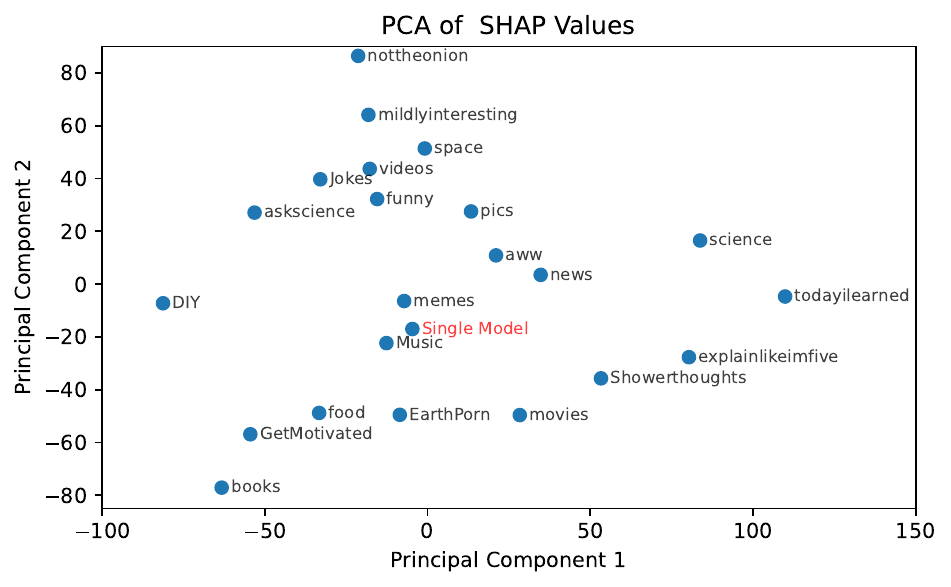}
    \caption{PCA of the SHAP values of the Single Model (Random Forest classifier) trained on the data from all the mainstream subreddits versus the SHAP values of the models trained on each single mainstream subreddit.}
    \label{fig:b3}
\end{figure}

\subsection{Temporal Robustness Analysis}
\label{sec:temporal}
We next examine the stability of classification performance over time prior to the users' first engagement with \rconsp. 
Specifically, we evaluate classifier performance on subsets of user comments ordered both chronologically and by cumulative activity within mainstream communities. 
Temporal analysis shows a small but statistically significant increase in accuracy---2.9 percentage points between the first and last activity windows (Mann-Kendall $p=0.03$). 
In contrast, classification performance remains stable across cumulative activity thresholds (Mann-Kendall $p=0.22$). 
Overall, these results indicate that the ability to distinguish conspiracy-engaged users is largely consistent regardless of the amount or timing of their prior mainstream activity. 
This also indicates that users’ psycholinguistic patterns are relatively stable before engaging with conspiracy communities, consistent with the hypothesis that initial engagement is driven in part by self-selection rather than solely by exposure within the platform~\cite{imhoff2022tearing,spohr2017fake}.
We leave more details and the plot of these results in the \Cref{app:temporal}.

\subsection{Feature Importance Analysis}
To identify the psycholinguistic features that characterize conspiracy-engaged users, we perform a SHAP analysis on subreddit-specific models, computing each feature’s contribution to the classification task and comparing feature importance patterns across subreddits.
By measuring their similarity, we evaluate whether conspiratorial language manifests consistently across mainstream communities or adapts to the specific norms of each context.
As shown in Panel \textbf{B} of \Cref{fig:s1_features}, while some subreddits exhibit more similar feature importance patterns, we do not see clear separations across groups of communities, thus indicating substantial variation across different contexts.
Additionally, in \Cref{fig:b3}, we show the results of applying Principal Component Analysis to the SHAP values of the Single Model, trained on data from all mainstream subreddits, as well as the models trained separately on each mainstream community, with all the users' activity thresholds aggregated. 
The figure illustrates that the Single Model is positioned near the origin ([0, 0]) in the PCA space, suggesting that it effectively represents an average of the community-specific models, which are more widely dispersed around the center.
This last analysis reinforces the idea that the language of conspiracy-engaged users is highly context-dependent, adapting to the norms of each community rather than reflecting a fixed linguistic pattern.

To further investigate these results, we qualitatively analyze specific clusters that emerged from our prior analyses. As shown in \Cref{fig:s1_features}, Panel B, the hierarchical clustering algorithm identifies five distinct clusters. For example, the yellow/gold cluster—comprising the subreddits r/DIY, r/GetMotivated, r/books, r/Music, and r/food—shares emo\_anger as a negative predictor. Specifically, a low presence of anger is associated with the non-conspiracy class, likely reflecting a tranquil environment where typical users discuss their hobbies, whereas conspiracy-leaning users display more anger in these spaces. 
Another notable example is the green cluster, which consists of r/science, r/todayilearned, r/news, r/Showerthoughts, and r/explainlikeimfive. 
Within this group, mental serves as a primary discriminant predictor, indicating a strong presence of reasoning and cognitive processes. 
This aligns with the high volume of explanatory language typical of these communities, making them the most likely venues for serious, real-world themes.

Overall, results from our study provide evidence for the existence of a large linguistic difference between conspiracy-engaged users and non-conspiracy-engaged users that can be captured in more mainstream spaces, also compared to other topic-adjacent subreddits.
However, they also suggest that there are no universal psycholinguistic fingerprints for conspiracy users discussing mainstream topics, as their manifestation varies across different mainstream and conventional communities and is highly context-dependent.
Nevertheless, we find recurring patterns such as negative emotions~\cite{korenvcic2024distinguishes} (anxiety, anger) and negative themes (death, illnesses), as shown in \Cref{tab:summary_feat}.
We report the top-10 most discriminant features for each mainstream subreddit we considered in our experiments in \Cref{fig:shap_1,fig:shap_2,fig:shap_3}. %\Cref{app:shap}.

\begin{table*}[]
\small
\centering
\caption{Examples of LIWC psycholinguistic features that appear with high, moderate, or low frequency as top discriminant features in the models trained on single subreddits. We also report the impact of that feature on the classification output. Positive: High values of the feature drive the classification towards conspiracy.  Negative: High values of the feature drive the classification towards non-conspiracy. Mixed: The features have different roles for different models in which it appears.}
\label{tab:summary_feat}
\begin{tabular}{llll}
\toprule
\textbf{Frequency} \hspace{2em} & \textbf{LIWC Feature} \hspace{2em} & \textbf{Impact} \hspace{2em} & \textbf{Subreddit Examples} \\
\midrule
	\textbf{$\ge$ 11 models} & filler                & Positive                & explainlikeimfive, EarthPorn, todayilearned \\
                         & WC (Word Count)       & Mixed        & askscience, science, mildlyinteresting \\
                         & sexual                & Mixed                & askscience, Music, books                    \\
                         & swear                 & Positive                & explainlikeimfive, science, DIY            \\
                         \cmidrule{1-4}
	\textbf{4 to 10 models}  & illness               & Mixed                & news, explainlikeimfive, pics               \\
                         & death                 & Mixed        & funny, mildlyinteresting, pics              \\
                         & emo\_anger            & Positive        & food, DIY, GetMotivated                     \\
                         & emo\_anx              & Mixed                & EarthPorn, movies, memes                    \\
                         & mental                & Mixed                & todayilearned, news, space                  \\
                         \cmidrule{1-4}
	\textbf{$\le 3$ models}         & curiosity             & Negative        & jokes                                      \\
                         & achieve               & Positive        & nottheonion                                 \\
                         & Clout            & Positive        & books                               \\
\bottomrule
\end{tabular}
\end{table*}

 \section{Discussion}\label{Discussion}
This work shows that there are detectable psycholinguistic differences between conspiracy-engaged and non-conspiracy-engaged users that are captured within mainstream communities. 
Our results align with previous research~\cite{klein2019pathways}, suggesting that users who engage with \rconsp display distinct linguistic patterns. 
While previous work identified these differences within communities already shaped by the activity of conspiracy theorists, we instead observe that such linguistic discrepancies emerge across various mainstream communities, with notable variations depending on the specific community. 
This result underscores both the complexity of this task and the critical role of social context in shaping the language of self-expression of individuals~\cite{Danescu2011www,danescu2013no}. 
Despite this variability, these linguistic fingerprints hold strong predictive power, enabling machine-learned models to achieve high classification accuracy, as shown in \Cref{fig:s1_both}.

These findings have important implications for the design of content moderation systems and the methodology of social media analysis. 
Specifically, the contrast between the robust performance of subreddit-specific models and the degradation of the aggregate model challenges the feasibility of ``one-size-fits-all'' detection approaches. 
We observed that applying a single model across all communities resulted in a performance drop of up to 17 percentage points compared to local models. 
This indicates that effective detection cannot rely on a universal linguistic ``fingerprint''; rather, it requires context-aware strategies that account for how users adapt their linguistic register to the specific norms of the community they inhabit.

Furthermore, the pervasiveness of these signals impacts our understanding of ``neutral'' online spaces. 
We detected strong discriminatory markers even in innocuous, low-stakes environments such as \textit{r/aww}, \textit{r/food}, or \textit{r/DIY}.
This result challenges the assumption that ideological signaling is strictly compartmentalized within political echo chambers~\cite{garimella2018political,cinelli2021echo} or radicalized forums~\cite{calikus2025risk}.
Instead, our results suggest that mainstream, general-interest communities serve as heterogeneous mixing grounds where distinct linguistic identities remain visible and measurable, regardless of the topic being discussed.

From a methodological standpoint, this study establishes that local feature importance is superior to global feature aggregation for characterizing user groups. 
As illustrated by our PCA analysis of SHAP values, the single aggregate model essentially averages out community-specific nuances, obscuring the specific linguistic levers that distinguish these users in different contexts. 
Consequently, we suggest that future research in computational social science should adopt an ``ecological'' approach, characterizing user behavior relative to the specific discursive norms of the environment rather than seeking fixed, platform-wide behavioral signatures.

\section*{Ethical Implications}
The ethical implications of this work extend beyond content moderation and speak to longstanding tensions between freedom of expression, individual rights, and the protection of public discourse~\cite{scanlon1972theory}.
A key ethical risk raised by our findings is the potential misuse of linguistic profiling at the user level. 
Although our models are intended for population-level analysis, similar techniques could be repurposed to infer ideological or psychological traits of individual users in mainstream, non-political communities~\cite{colacrai2024navigating}.
Such inferences are inherently probabilistic and context-dependent, and misclassification may result in unjustified scrutiny, stigmatization, or harm, particularly if applied without transparency, consent, or avenues for contestation.
Moreover, the detectability of linguistic differences in general-interest communities increases the risk of preemptive or identity-based moderation strategies that target inferred user traits rather than observable harmful behavior. 
These approaches may produce negative effects on legitimate speech and disproportionately impact dissenting or minority viewpoints. 
For these reasons, the results presented here should not be used to justify early reactive interventions such as content removal or de-platforming. 
This caution is especially warranted given evidence that conspiracy theories are resilient to direct suppression and that heavy-handed enforcement may reinforce conspiratorial beliefs~\cite{monti_online_2023}.

\section*{Limitations}
Our study has some limitations that we acknowledge and discuss. 
The first limitation is the reliance on the psycholinguistic features provided by dictionary-based computational methods.
While these features are widely used in the literature~\cite{tadesse2019detection,faasse2016comparison,giachanou2023detection}, and the dictionary is based on well-established psychological theories, they may still fail to capture the full nuances of online discourse.
A dictionary-based approach primarily focuses on individual word frequencies, overlooking more complex linguistic structures such as sentence context, syntactic dependencies, and discourse-level features that could be crucial for distinguishing conspiratorial narratives from general discussions~\cite{garten2018dictionaries}.
Nonetheless, this tool was still able to provide meaningful insights into the language of online users, as the machine-learning models we employ in our studies achieve a very high prediction accuracy.
Future work could test whether more powerful transformer-based strategies further increase the performance of the classifiers. Applying an interpretable framework to attention weights could capture linguistic structures and nuances that a dictionary-based method, such as LIWC, cannot measure. We expect that these endeavors would not fundamentally alter our core conclusion: that linguistic signals associated with conspiracy engagement are context-dependent across communities.
Another limitation is that while our experimental design is based on a large-scale longitudinal dataset, it does not allow us to establish a causal relationship between the difference in user language and their engagement with the conspiracy community, nor is the directionality of this relation explored in this study.
Finally, our approach focuses on users' direct activity, specifically, their commenting behavior, rather than their passive exposure to conspiratorial content, which we are unable to measure given the lack of available data. 
While comments serve as an observable proxy for user engagement, this limitation may overlook more subtle forms of exposure or interaction, such as lurking, which describes most of the activity of Internet users~\cite{sun2014understanding}.
Moreover, our strategy for identifying conspiracy-engaged users relies on self-identification and thus overlooks the cases in which these beliefs are shared in other communities without the user self-identifying as conspiracy-engaged.
Nonetheless, despite this constraint, our method successfully identifies a substantial number of users who go on to participate in conspiracy communities, offering valuable insights into the behavioral patterns of conspiracy theorists based on their online self-expression.

%\bibliographystyle{acl_natbib.bst}
\bibliography{main}

\appendix

\section{Effect of engagement level in \rconsp on linguistic patterns}
\label{app:disagg}
\begin{figure}[t]
    \centering
    \includegraphics[width=1\linewidth]{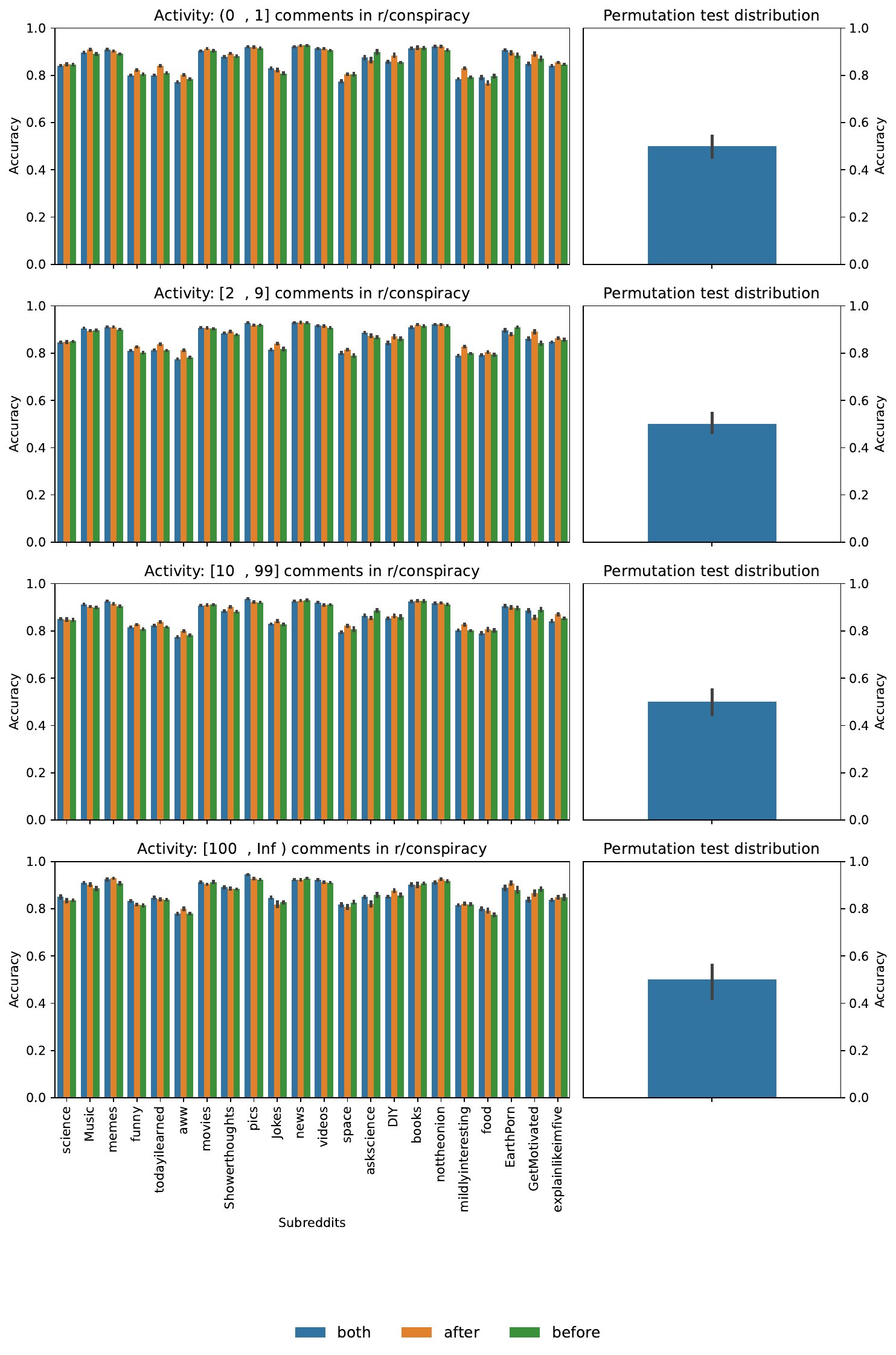}
    \caption{Disaggregated classification performance on the test set, together with the permutation tests performance for the four different classes of activity.}
    \label{fig:b1}
\end{figure}

We conduct a stratified analysis to examine how varying levels of activity in \rconsp influence model performance.
This analysis is motivated by one of our assumptions, which posits a potential link between the level of engagement in a conspiracy community and the degree of linguistic difference exhibited by an online user. 
To test this, we divide users active on \rconsp into four groups based on their number of comments: (0, 1], (1, 10), [10, 100), and [100, $+\infty$).
For each group, we construct a new dataset composed of the psycholinguistic feature vectors of conspiracy users within that group, along with a balanced random sample of non-conspiracy users. 
We then replicate the classification pipelines from the main, obtaining results consistent with those from the aggregated analyses. 
\Cref{fig:b1} shows the results of the experiments, disaggregated by the activity thresholds defined in \Cref{sec:Methods}. 
As noted, model performance is consistent across the different levels of user activity on \rconsp.
These results are supported by significant permutation tests ($p < 0.01$) across all activity classes and experimental conditions.

\subsection{Avoiding bias}
\label{A1}
\begin{figure}[t]
    \centering
    \includegraphics[width=1\linewidth]{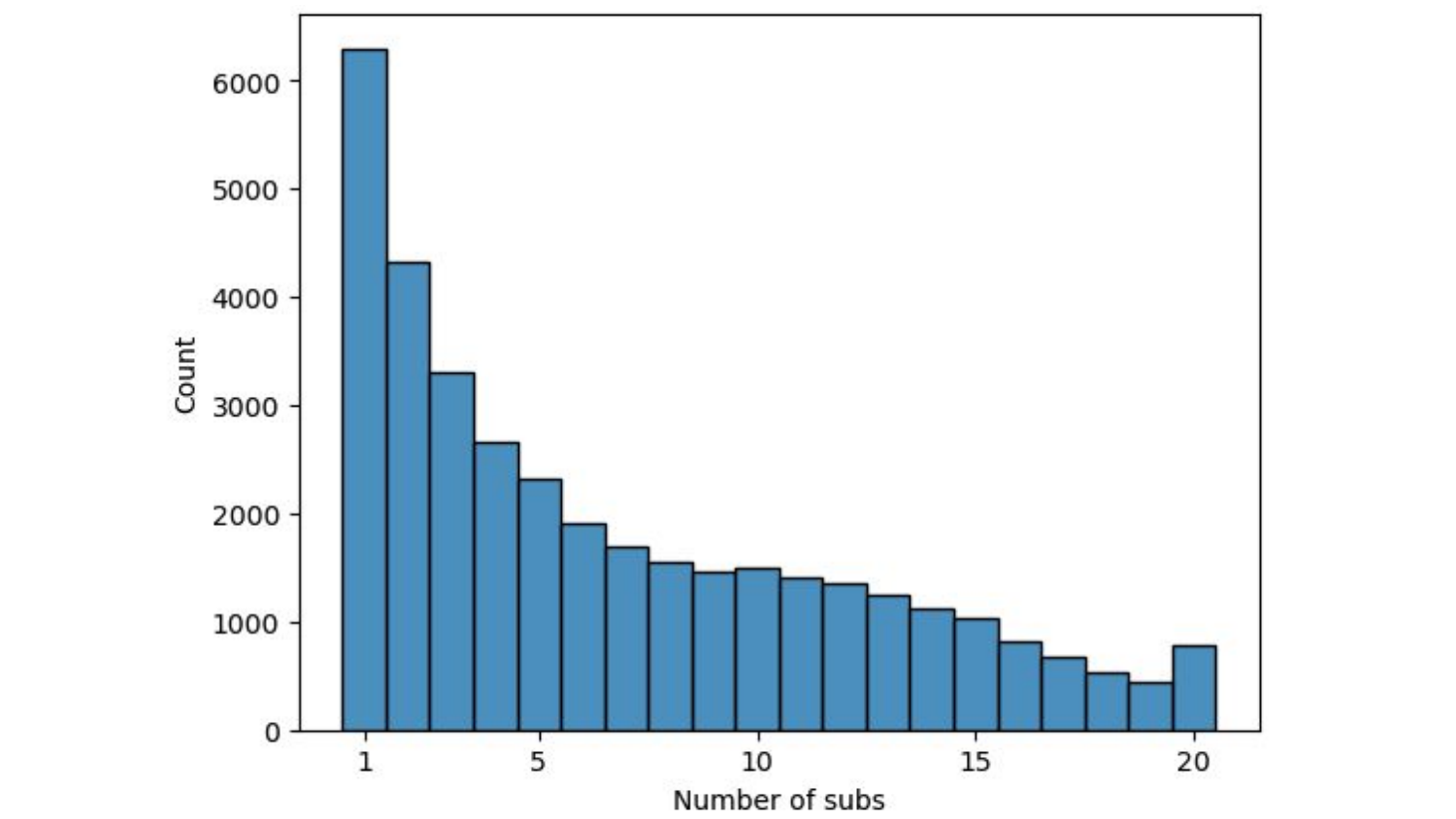}
      \caption{Example of bias in the selection of subreddits using the activity of conspiracy users as a proxy. The users we consider are those with at least 100 comments on \rconsp.}
    \label{fig:b6}
\end{figure}

\Cref{fig:b6} shows the result of a preliminary experiment we performed to select the non-conspiracy subreddit to which to focus our analyses.
We found that if we employed the activity outside of \rconsp of conspiracy users as a driver for this search, we would incur bias, as we would be selecting subreddits that would not correctly represent the pattern of activity of normal Reddit users, as shown in the figure.

\subsection{Additional Socio Demographics Analysis}
\label{app:sociodemo}
\begin{figure}[t]
    \centering
    \includegraphics[width=1\linewidth]{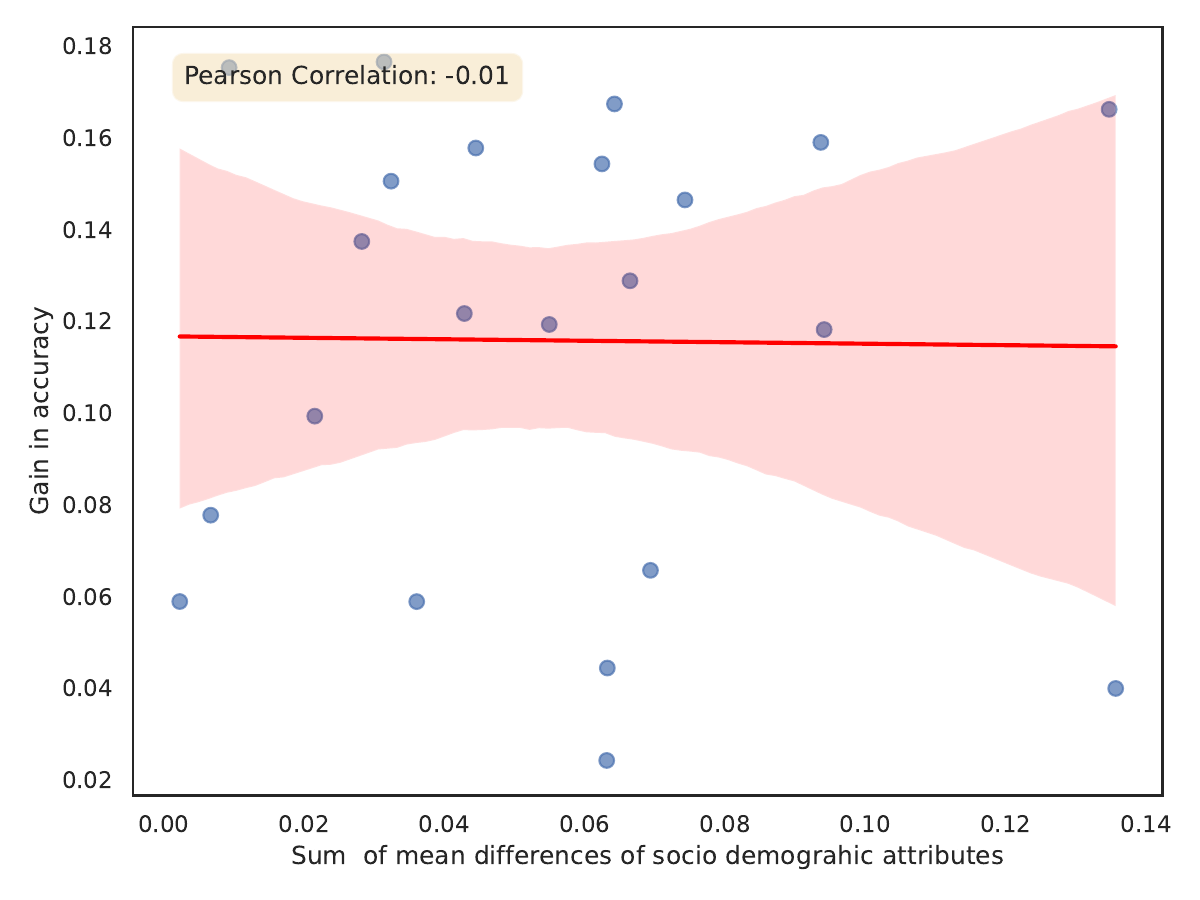}
    \caption{Correlation between the gain in accuracy of the subreddit-specific classifier on the general classifier and the absolute difference between the conspiracy users' attribute distribution and the given subreddit embedding.}
    \label{fig:b_11}
\end{figure}

As described in \Cref{sec:Methods}, we are interested in measuring the interplay of socio-demographics and accuracy of the classification.
To do so, we represented each user by computing a weighted average of the embeddings of all subreddits in which they were active, where the weights correspond to the relative frequency of their contributions across subreddits.
This procedure yielded a vector representation for each user, intended to capture socio-demographic and cultural orientations encoded in subreddit participation patterns.
To evaluate whether these embeddings introduced systematic bias into our analyses, we compared the distribution of conspiracy-engaged users active within a given mainstream subreddit to the embedding of that same subreddit.
Next, we quantified the absolute difference between the mean vector of conspiracy users and the embedding of each subreddit. 
We then correlated this measure with the improvement in classification accuracy obtained by the subreddit-specific model (described by \Cref{fig:s1_features} in the Results section).
The results, shown in \Cref{fig:b_11}, revealed no meaningful correlation, suggesting that socio-demographic features, as captured by subreddit embeddings, do not significantly account for the performance gains of our classifier. 
This provides empirical support for the decision to exclude socio-demographic attributes from the core design of our study.

\begin{figure}
    \centering
    \includegraphics[width=1\linewidth]{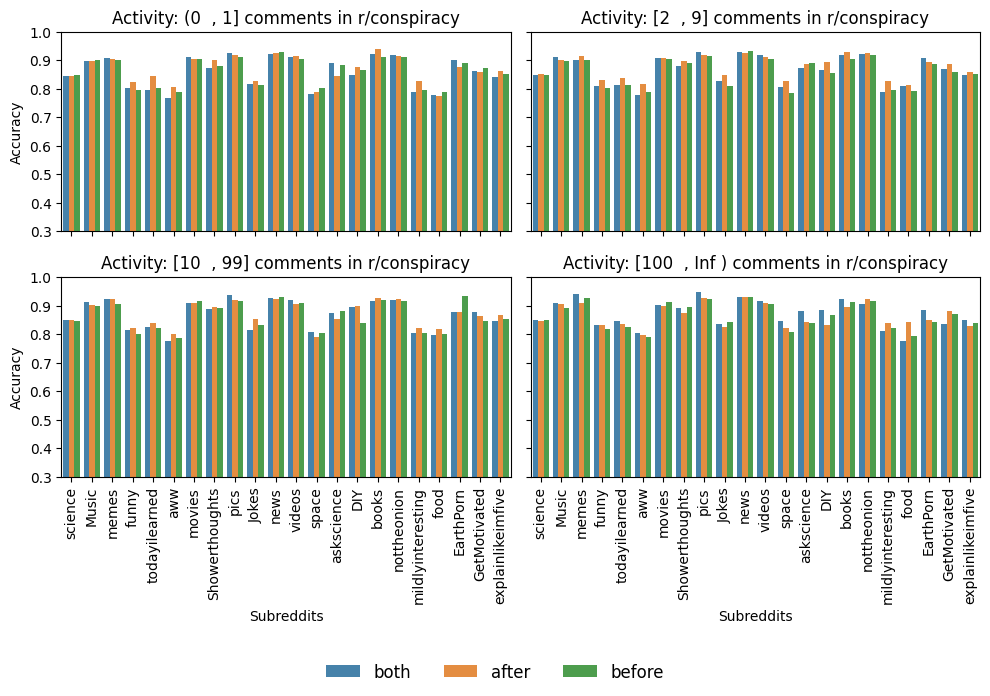}
    \caption{Disaggregated TabPFN experimental results.}
    \label{fig:b5}
\end{figure}

\subsection{Temporal Robustness Check}
As introduced in \Cref{sec:temporal}, we perform additional robustness checks on the classification results we obtained in our experiments by segmenting the users' activity into quintiles, based on two criteria: the temporal order and the cumulative order.
Meaning, we bin the comments of users into five different groups based on the moment in time and in which order they were posted by the user during their activity on Reddit prior to their first comment on \rconsp.

We apply this framework to each mainstream subreddit in our collection and for each group of users with varying levels of activity on \rconsp.
The datasets generated from these classification setups are balanced, and we employ them to train and evaluate a series of Random Forest classifiers, one for each top subreddit in our collection.
We then repeat the same operations we described in the previous analyses, i.e., feature normalization, hyperparameter search, and model training.
\label{app:temporal}
\begin{figure*}[!t]
    \centering
    \includegraphics[width=\linewidth]{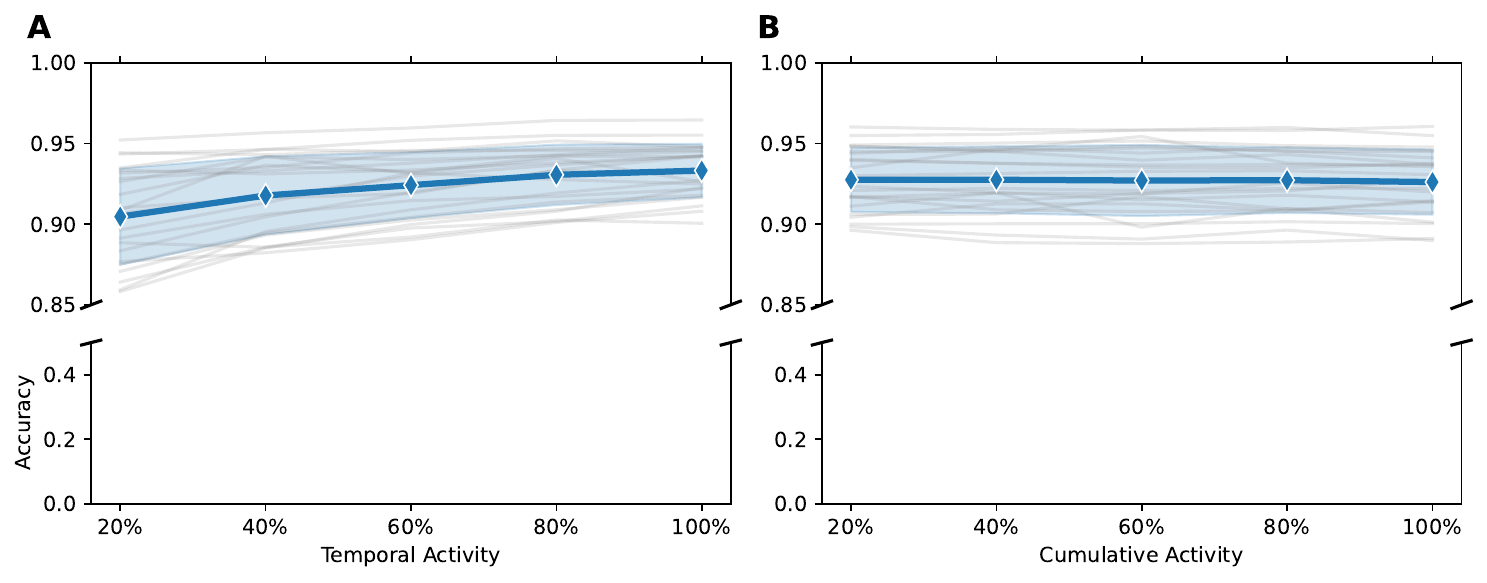}
    \caption{\textbf{Results of additional classification experiments on temporal robustness.} \textbf{A}: Classification accuracy across different temporal activity windows (based on comment order).
\textbf{B}: Classification accuracy across different numerical activity windows (based on cumulative comment volume).
In both cases, accuracy remains stable across thresholds, indicating no significant linguistic shift over time.}
    \label{fig:s2_curves}
\end{figure*}

% \subsection{Controlling for socio-demographic attributes}
% \label{app:sociodemo}
% In \Cref{fig:b_10,fig:b_11} we show the results of the process to control for socio-demographic features described in \Cref{sec:Methods}.
% We see how the socio-demographic score distribution of the conspiracy user is sometimes different from the average score of the mainstream subreddit and how this does not correlate with the gain in performance of the subreddit-specific classifier over the general classifier.

\subsection{Top features for each mainstream subreddit}
\label{app:shap}

\begin{figure*}
    \centering
    \includegraphics[width=0.45\linewidth]{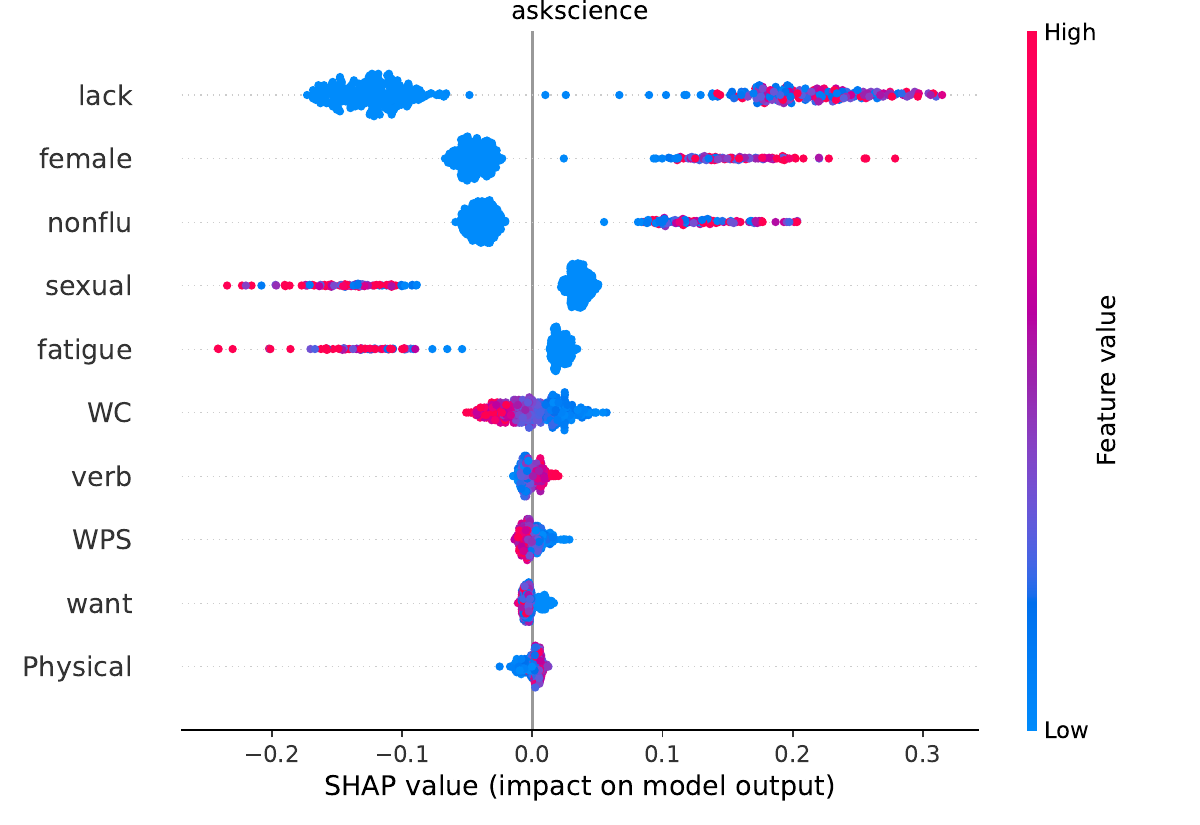}
     \includegraphics[width=0.45\linewidth]{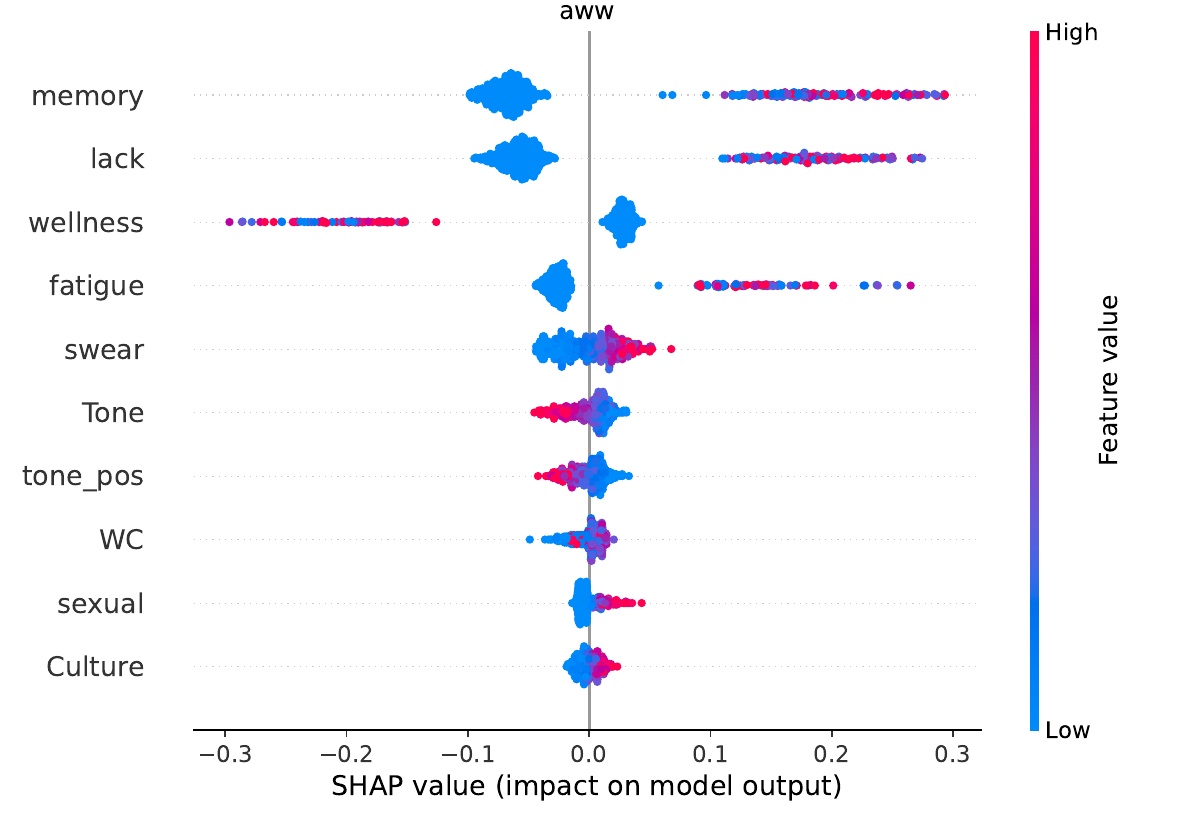}

    \includegraphics[width=0.45\linewidth]{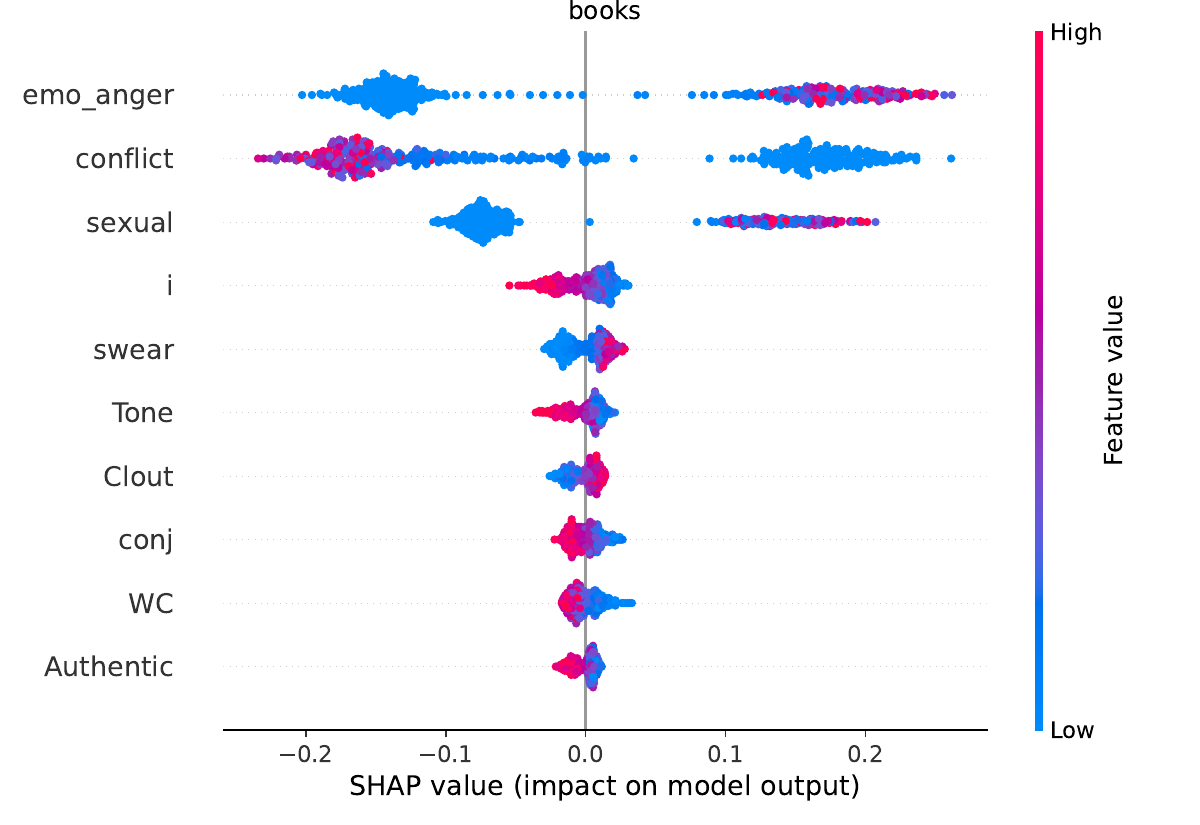}
    \includegraphics[width=0.45\linewidth]{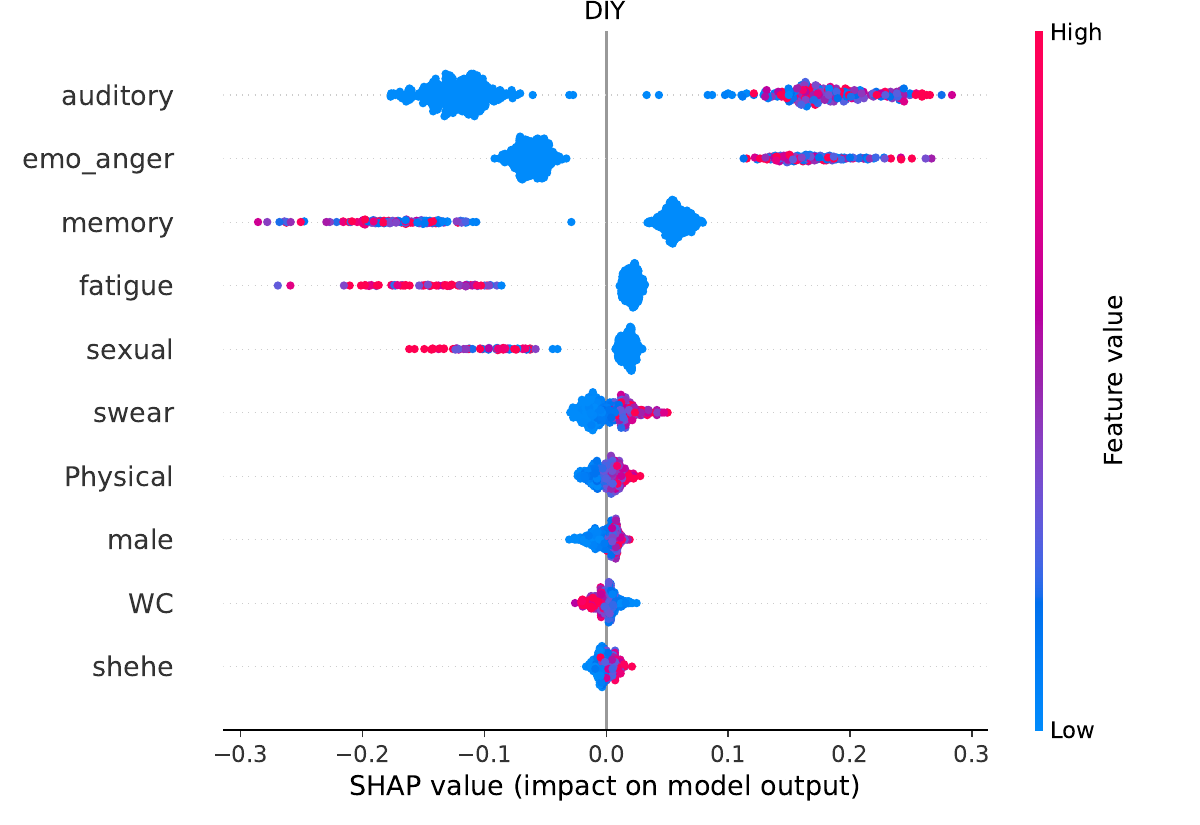}

    \includegraphics[width=0.45\linewidth]{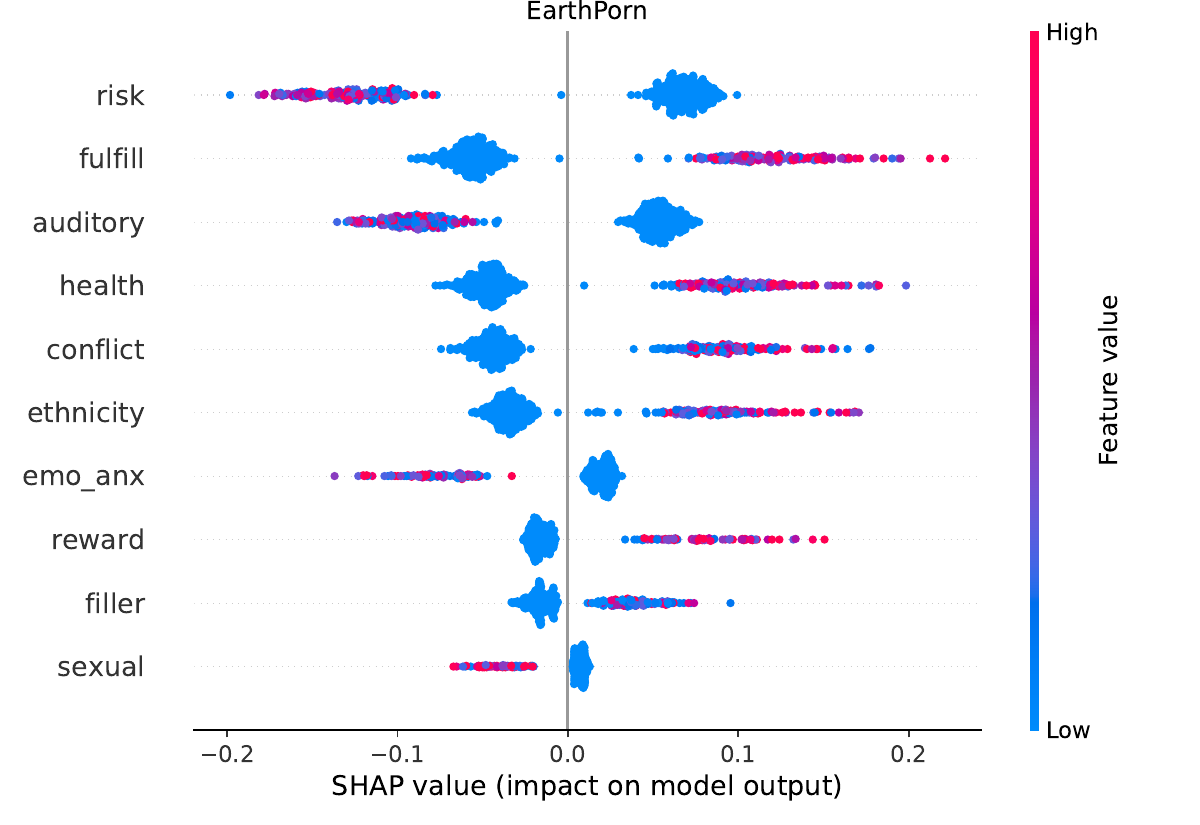}
    \includegraphics[width=0.45\linewidth]{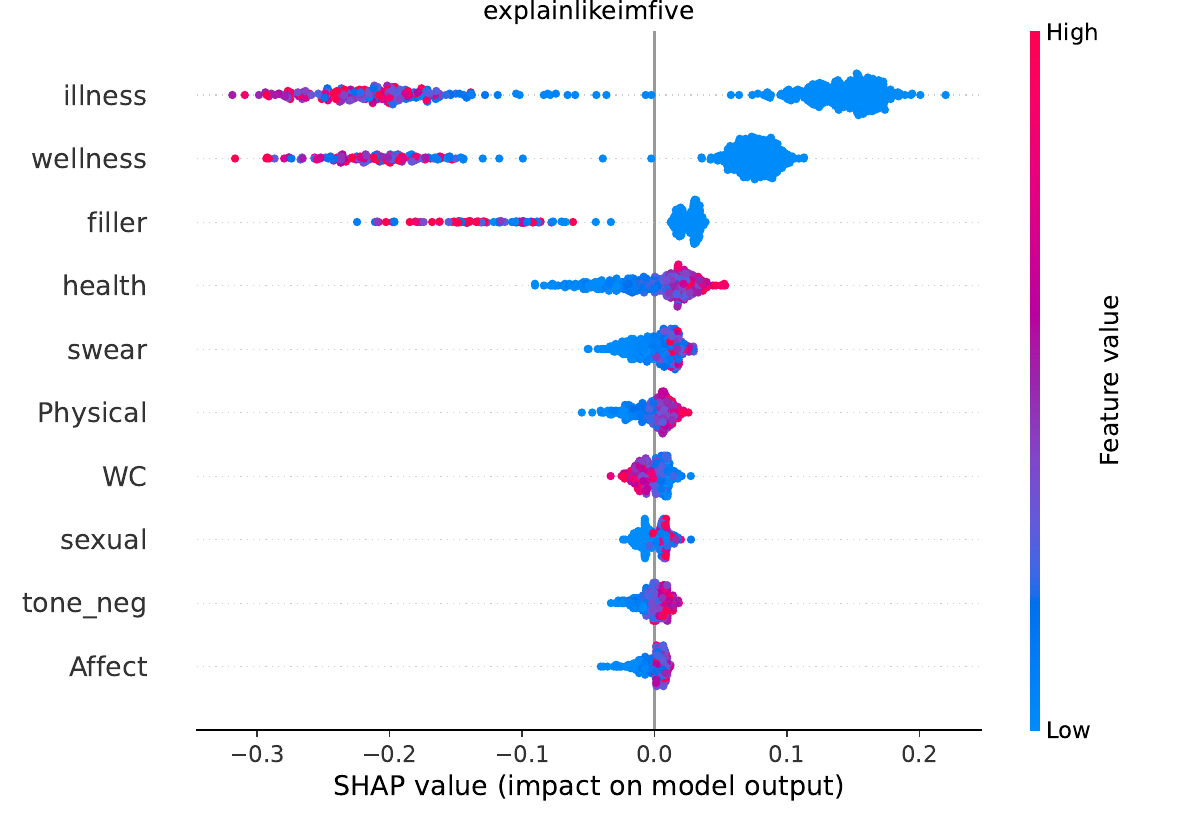}

    \includegraphics[width=0.45\linewidth]{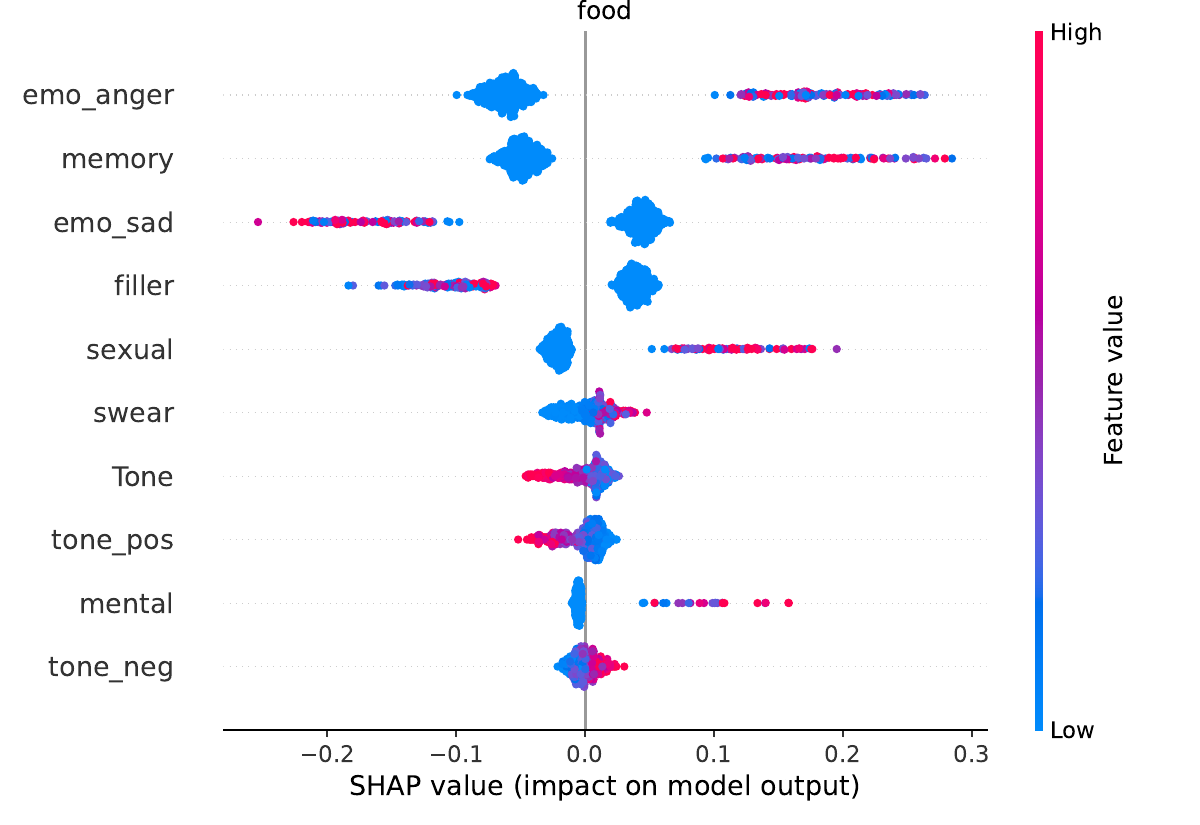}
    \includegraphics[width=0.45\linewidth]{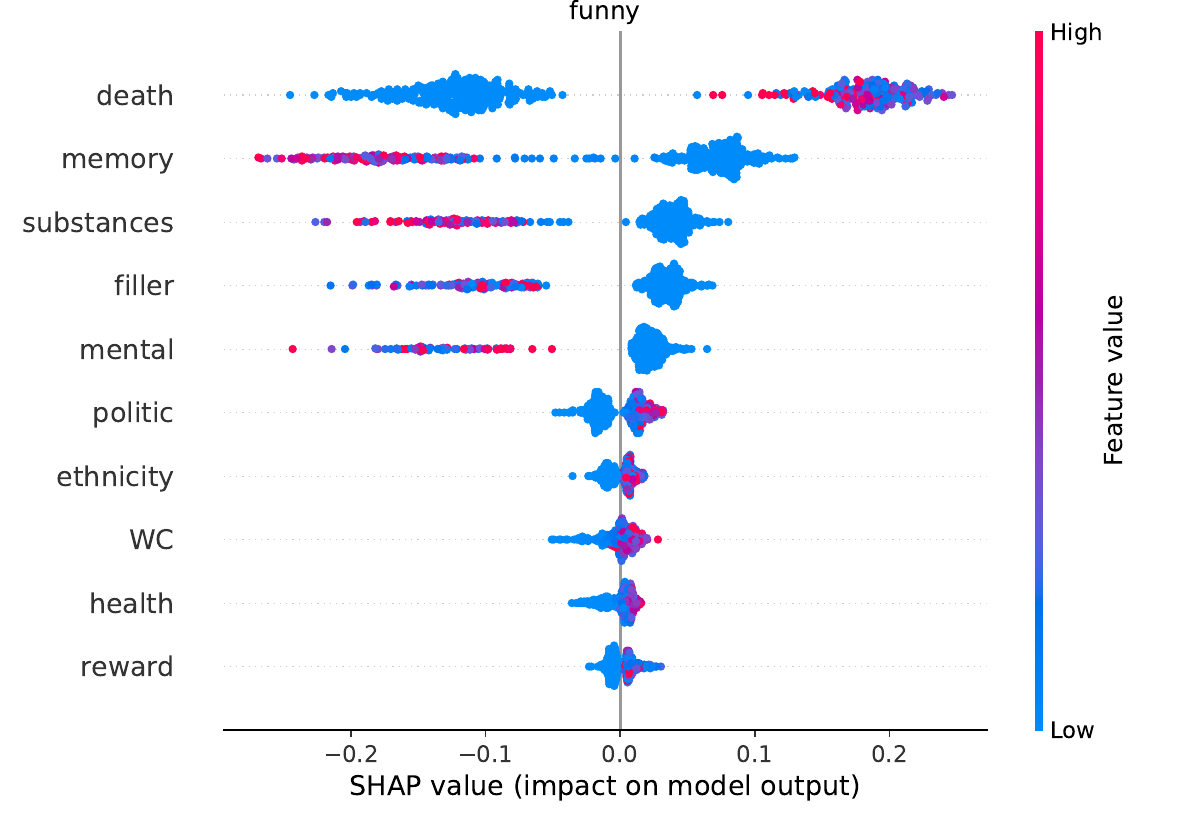}
    \label{fig:shap_1}
\end{figure*}

\begin{figure*}
    \centering
    \includegraphics[width=0.45\linewidth]{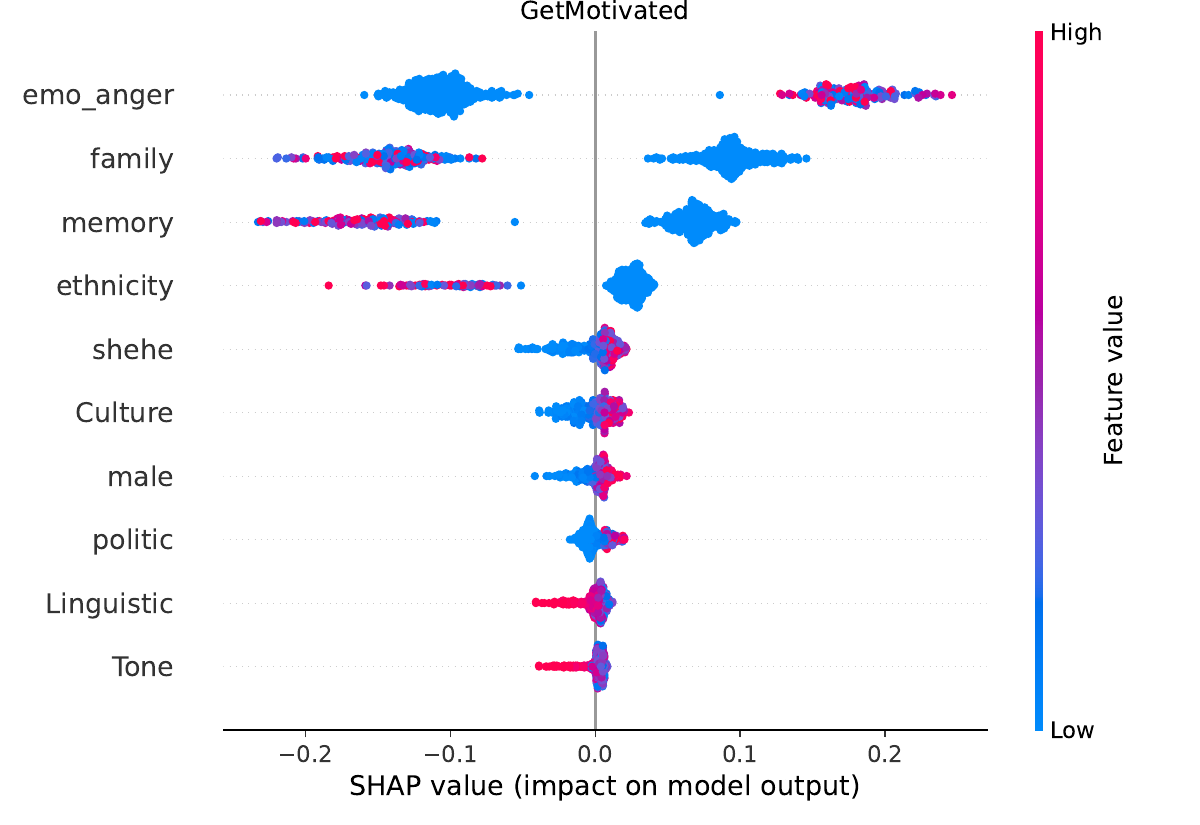}
    \includegraphics[width=0.45\linewidth]{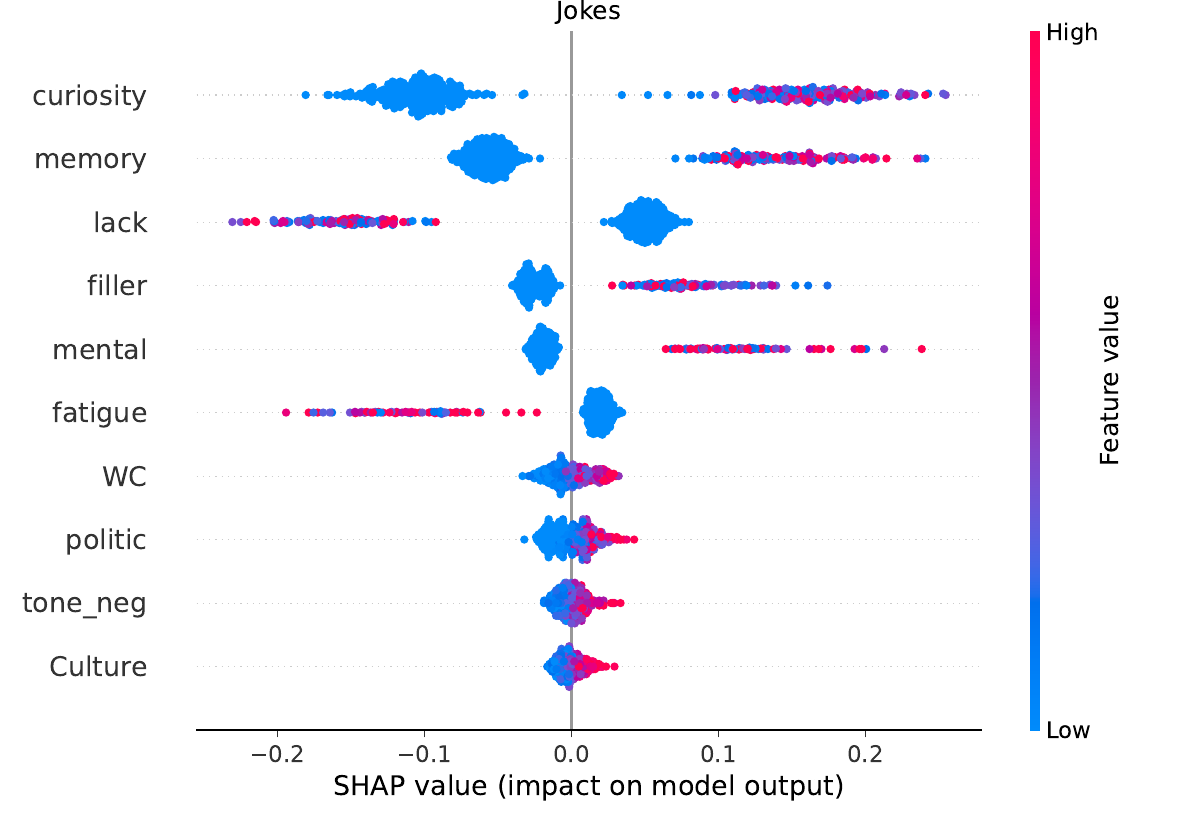}

    \includegraphics[width=0.45\linewidth]{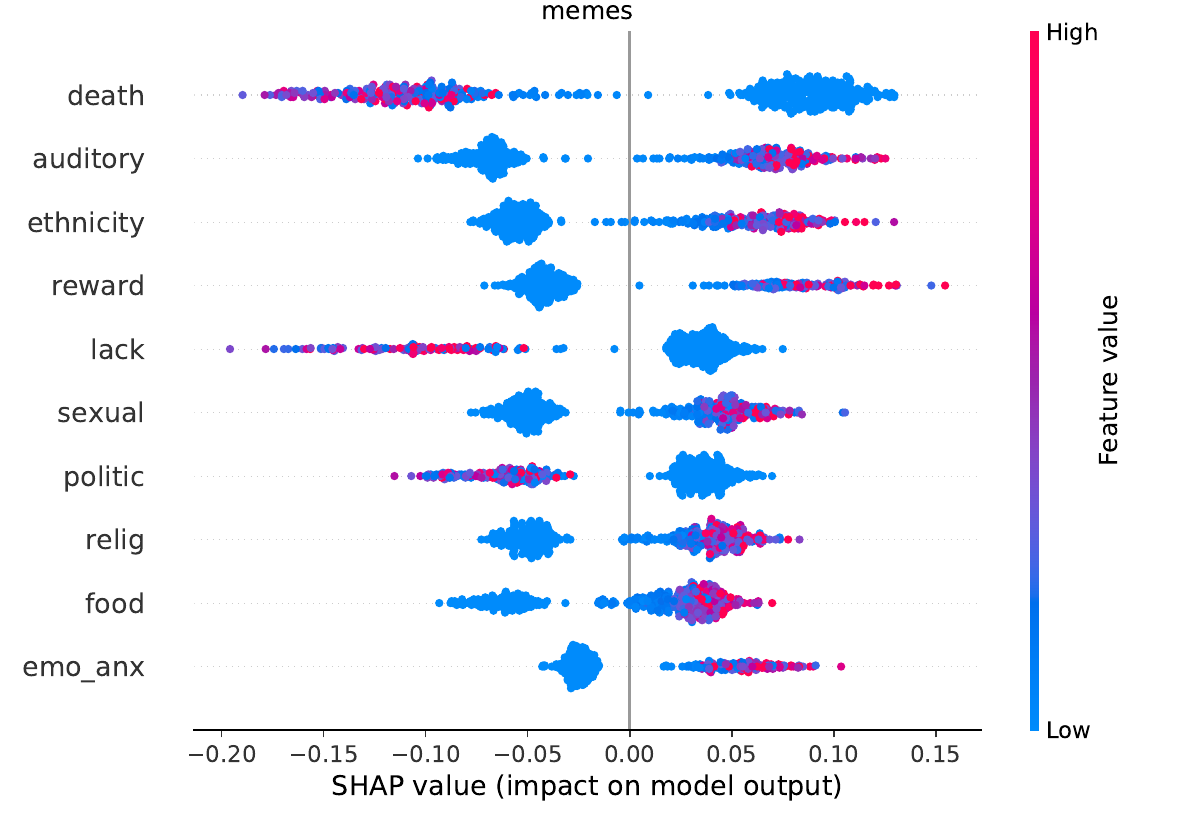}
    \includegraphics[width=0.45\linewidth]{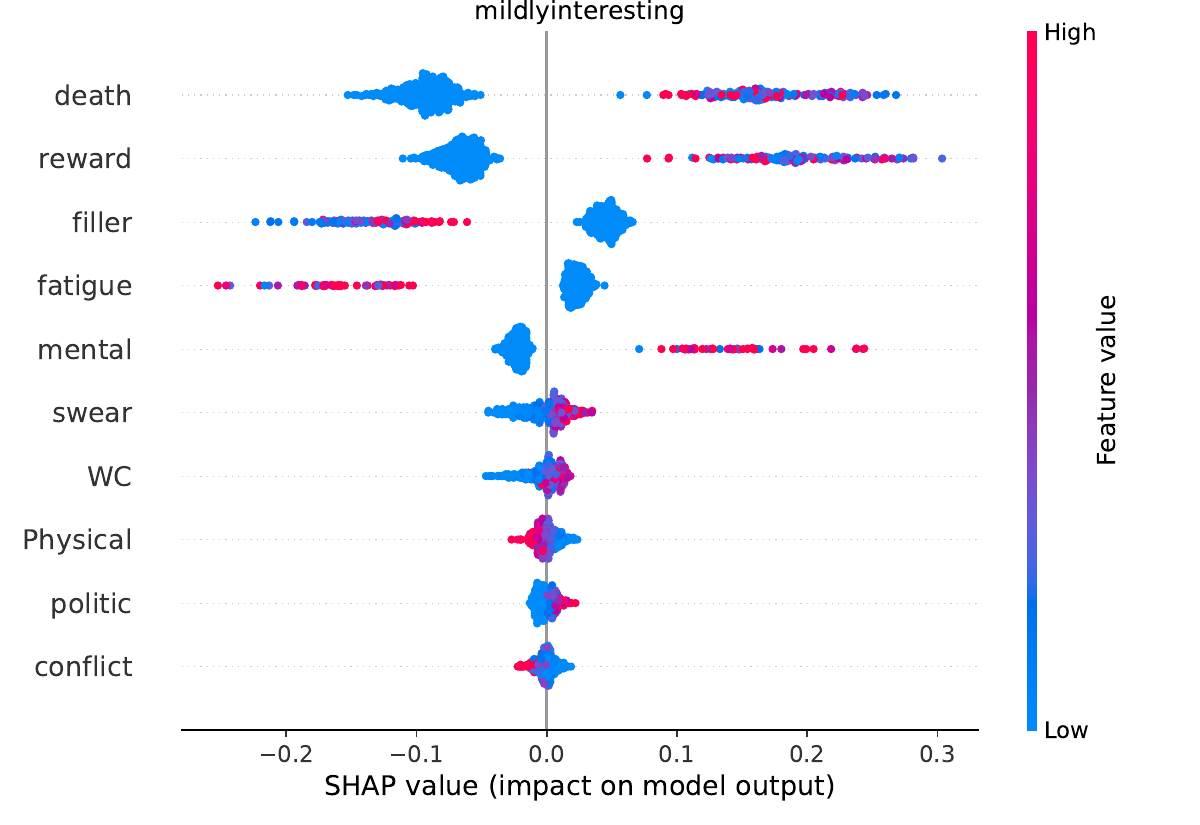}
    
    \includegraphics[width=0.45\linewidth]{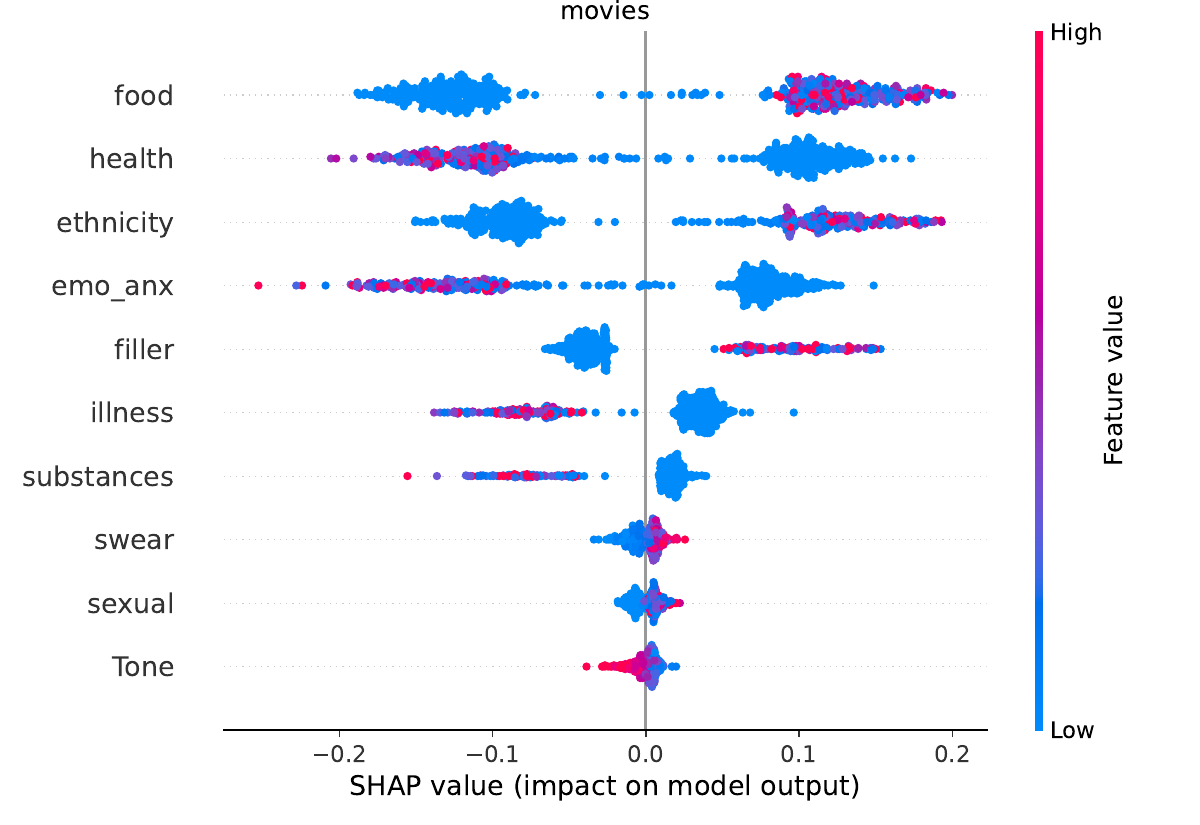}
    \includegraphics[width=0.45\linewidth]{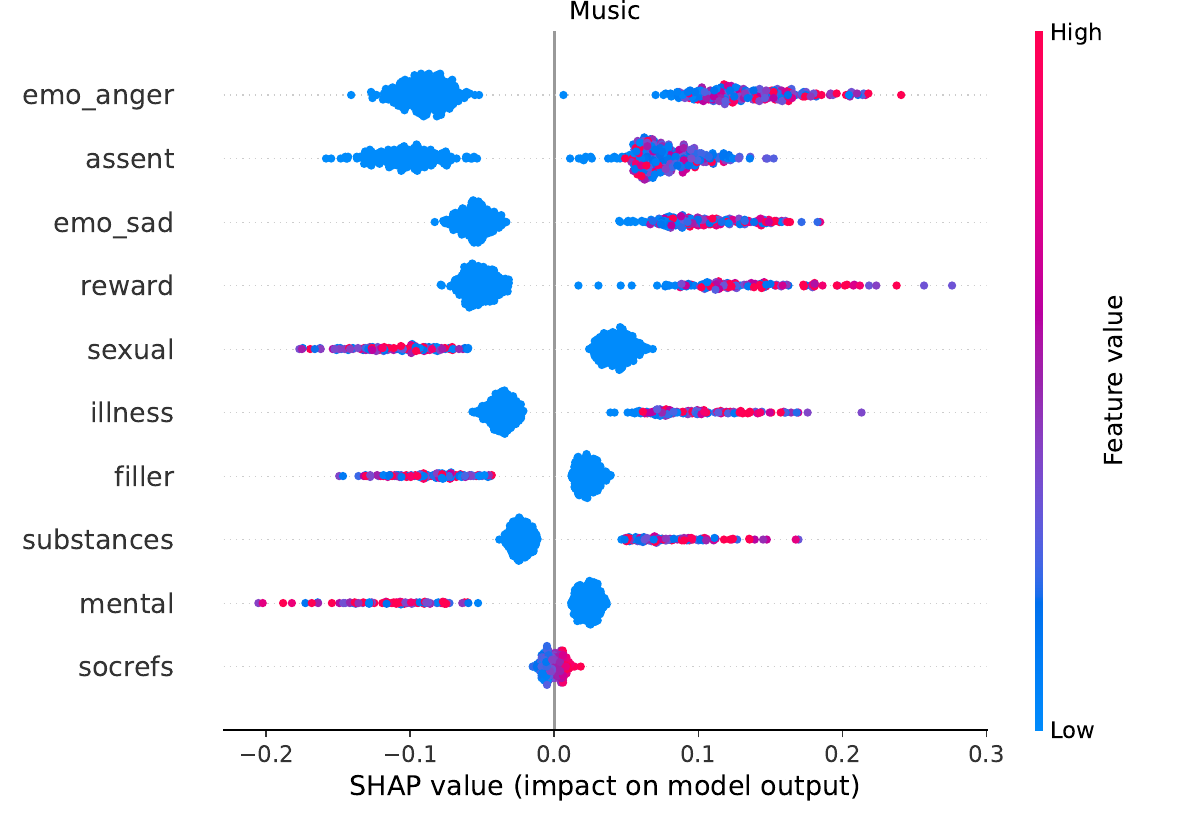}

    \includegraphics[width=0.45\linewidth]{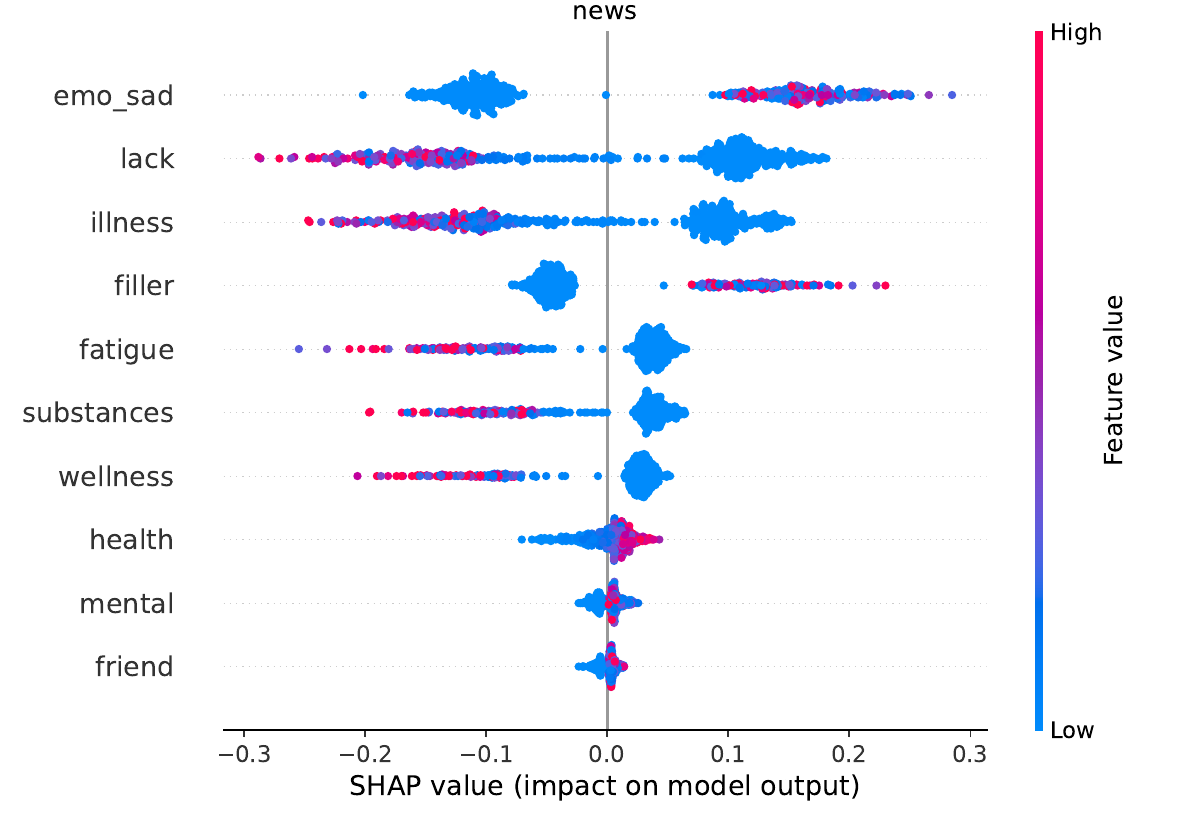}
    \includegraphics[width=0.45\linewidth]{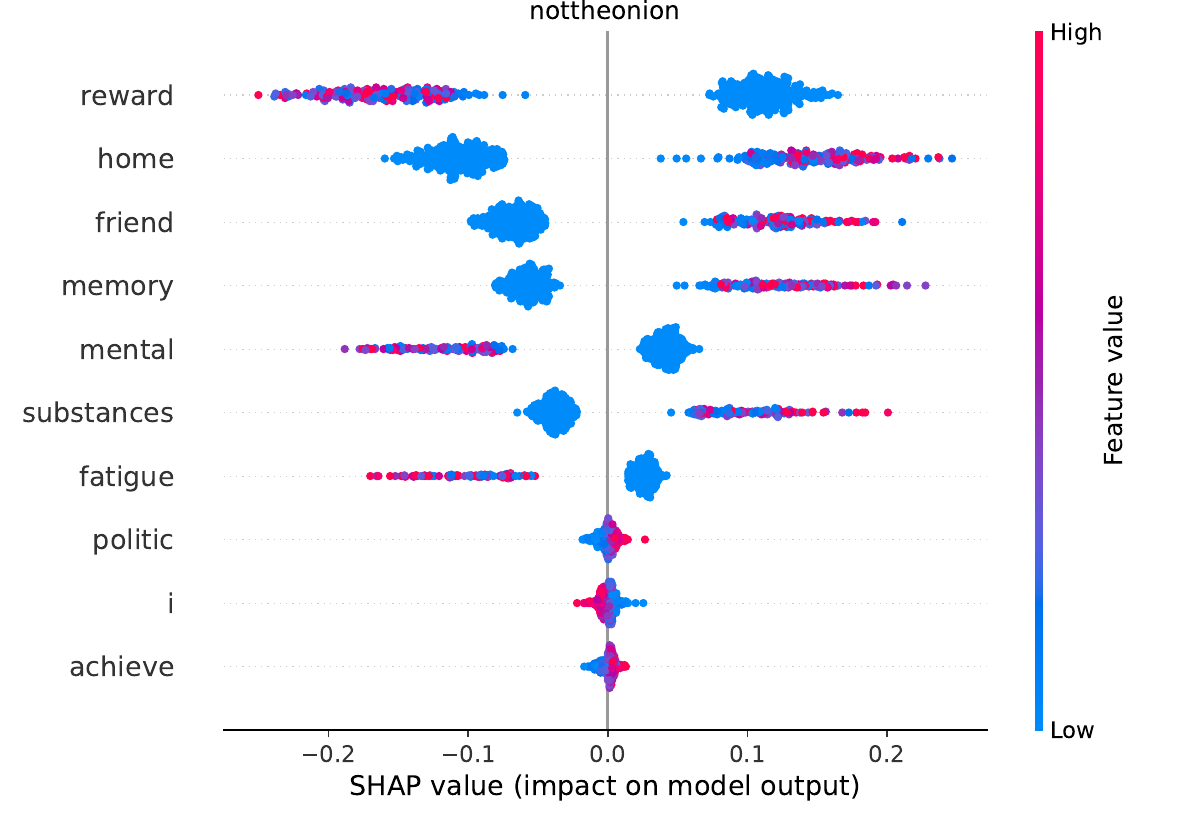}
    \label{fig:shap_2}
\end{figure*}

\begin{figure*}
    \centering
    \includegraphics[width=0.45\linewidth]{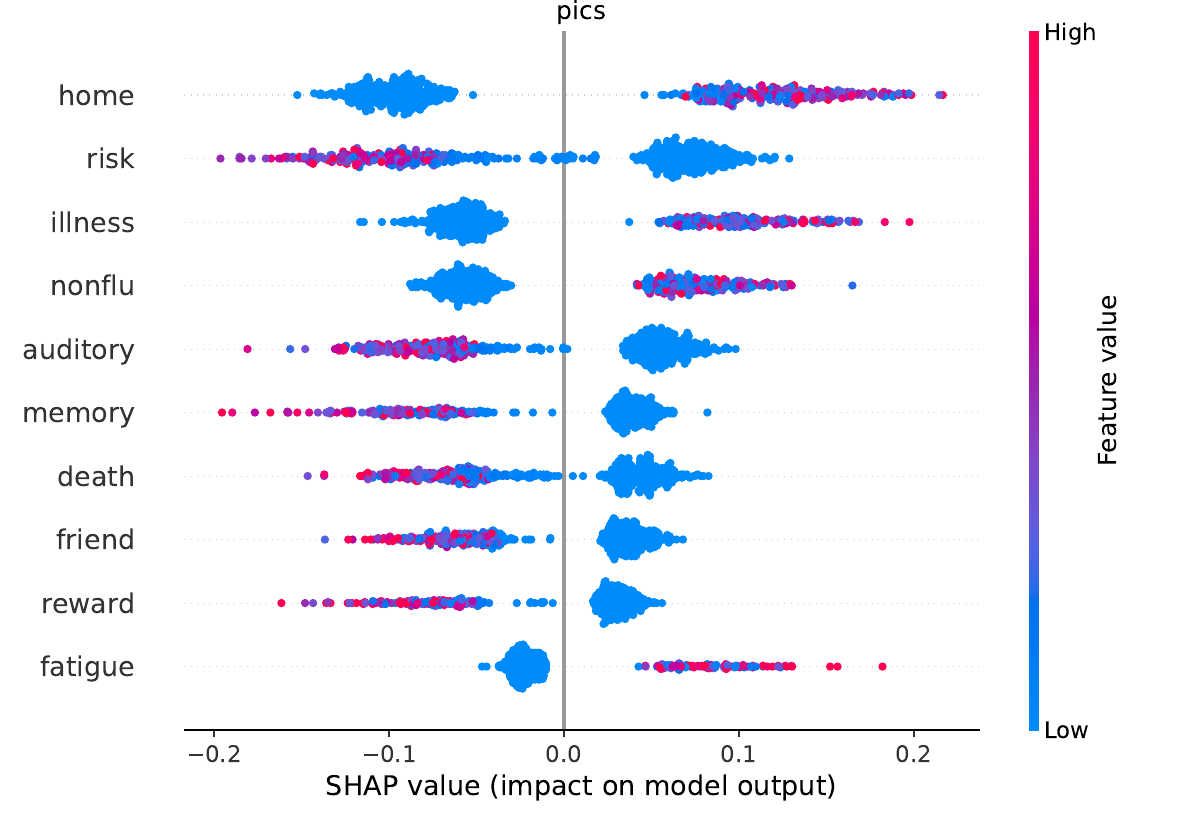}
    \includegraphics[width=0.45\linewidth]{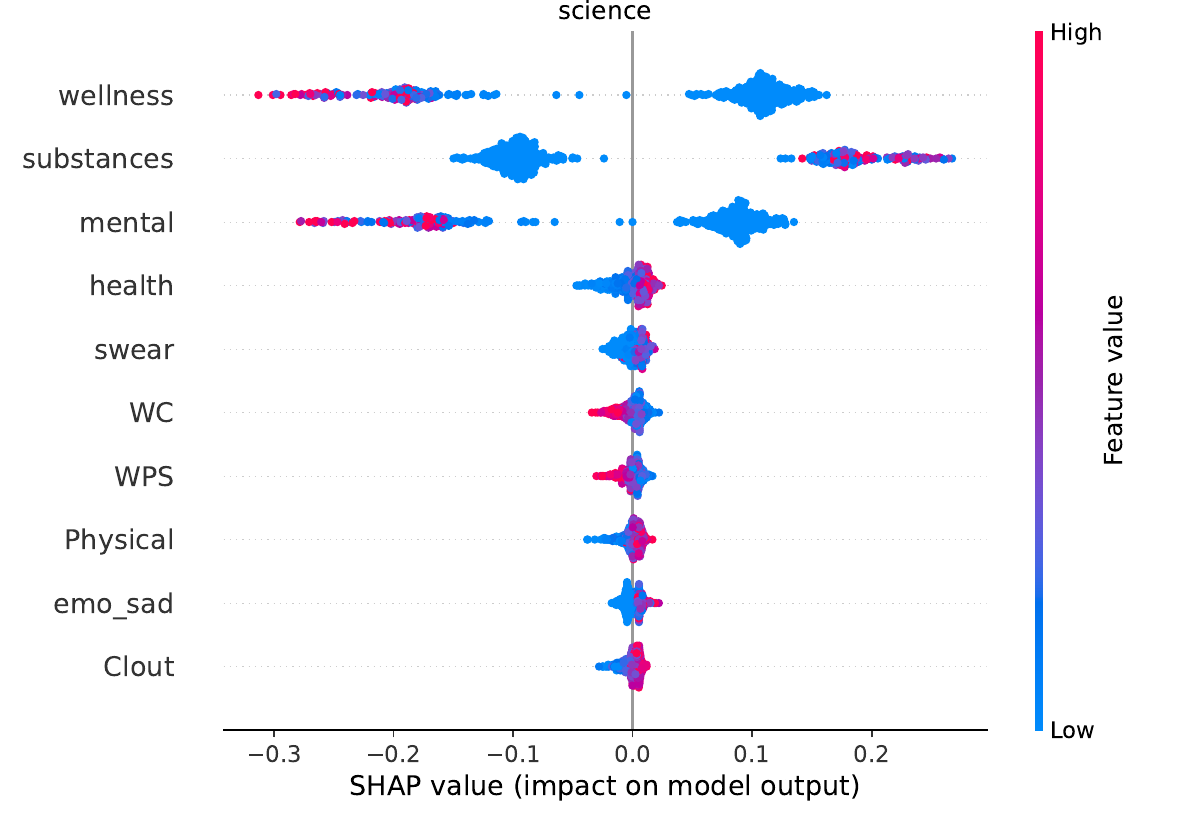}

    \includegraphics[width=0.45\linewidth]{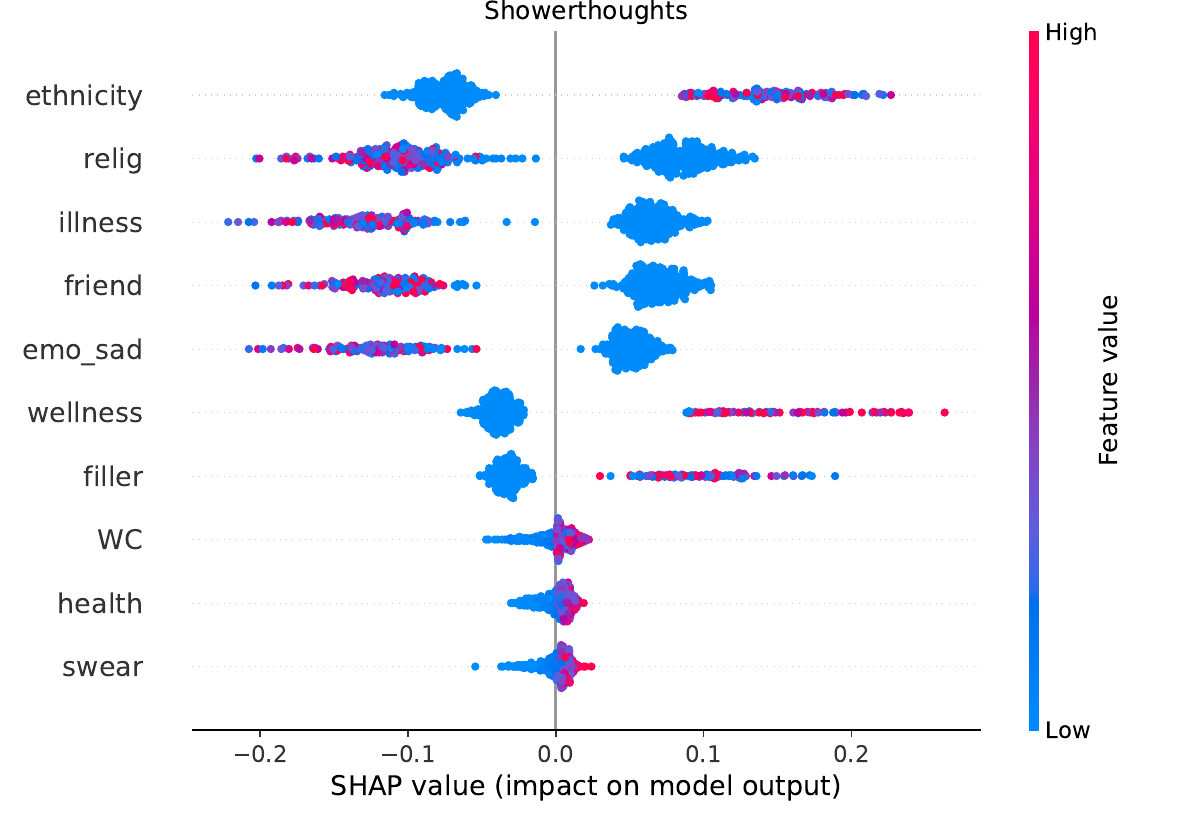}
    \includegraphics[width=0.45\linewidth]{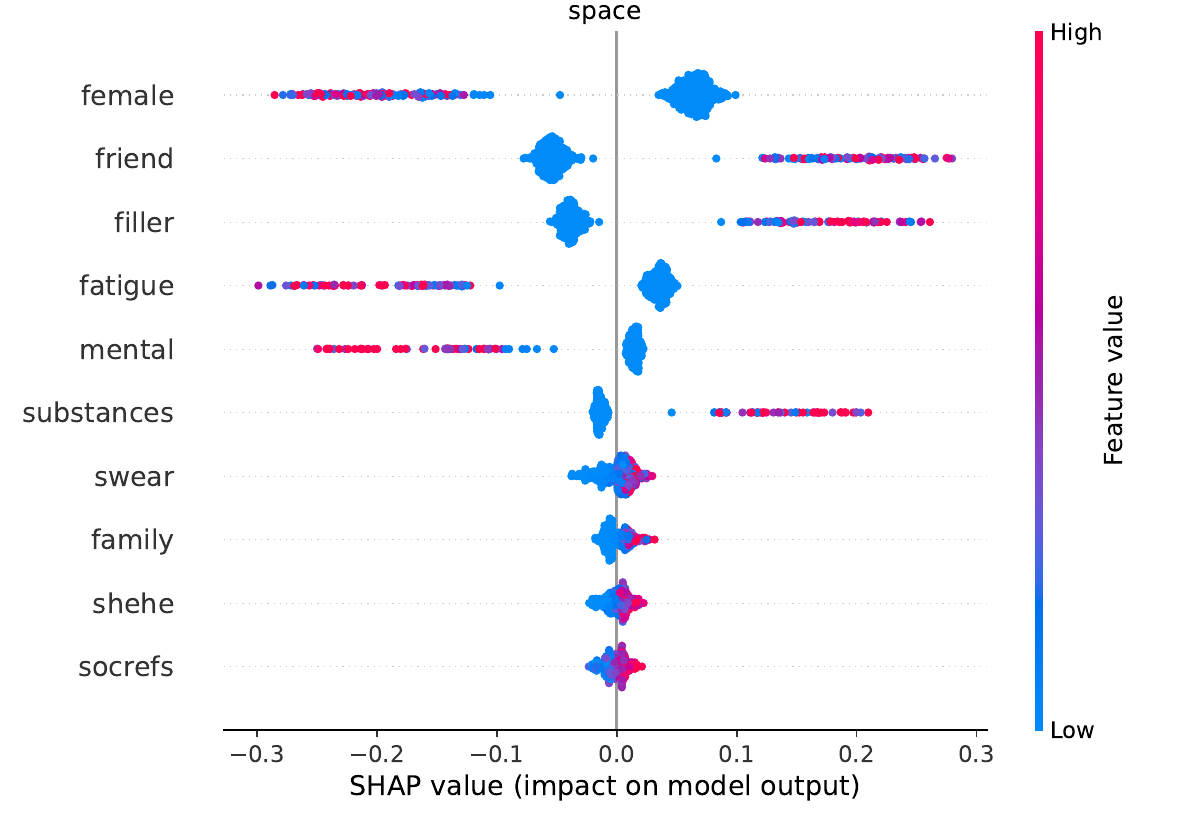}
    
    \includegraphics[width=0.45\linewidth]{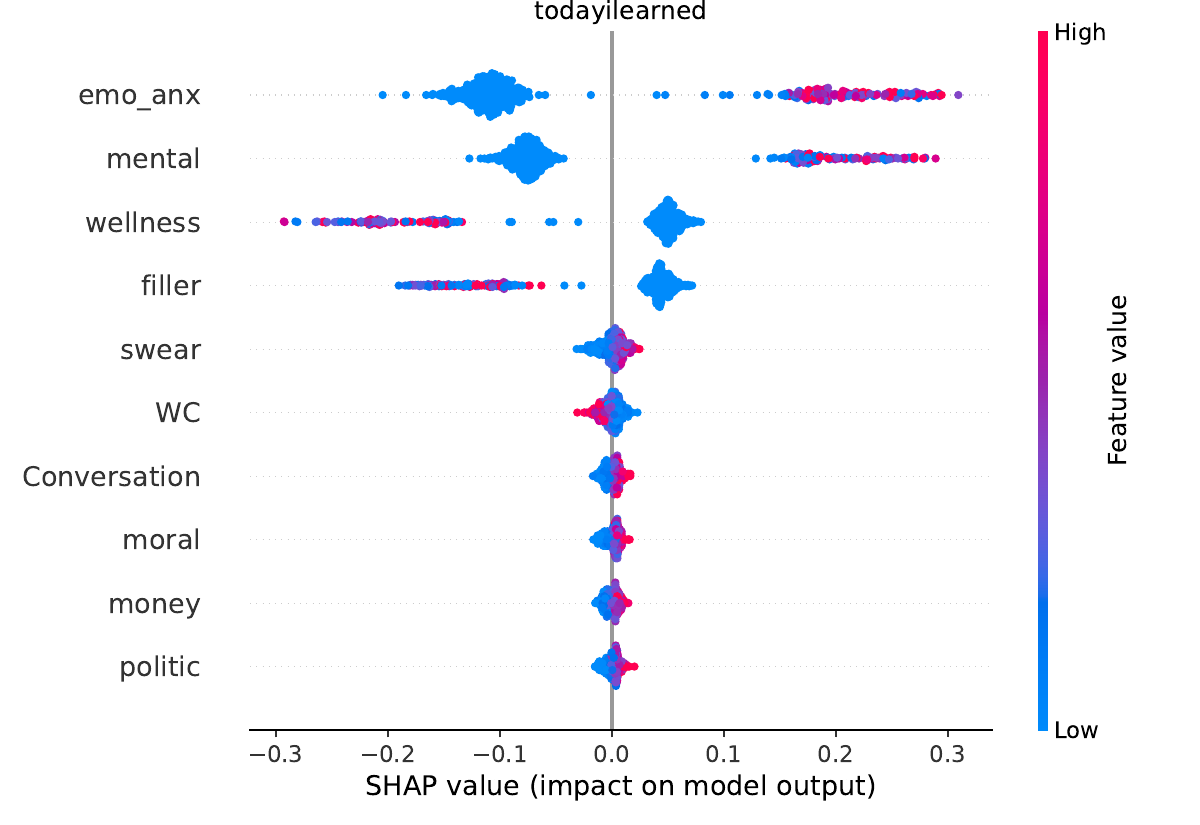}
    \includegraphics[width=0.45\linewidth]{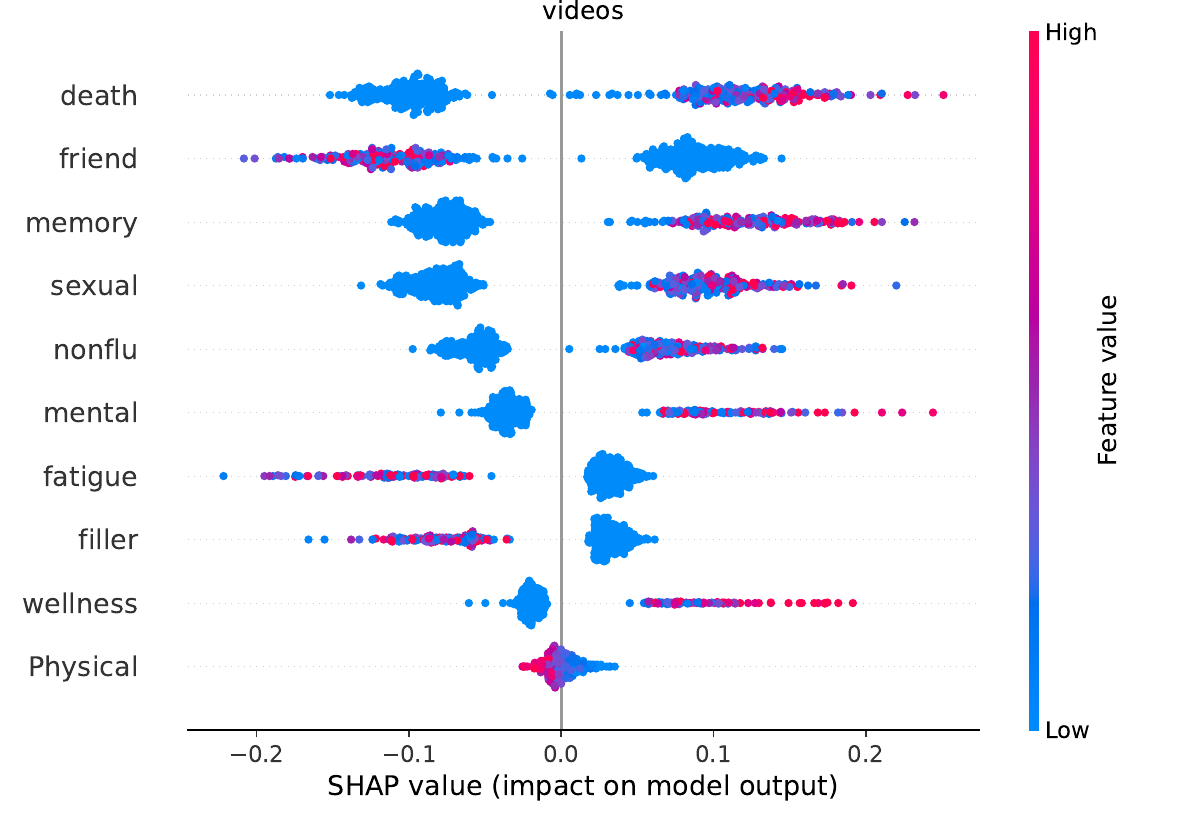}
    \label{fig:shap_3}
\end{figure*}

In \Cref{fig:shap_1,fig:shap_2,fig:shap_3} we show the top 10 most discriminant features in the decision tree together with the relative shap value, indicating the contribution of that feature, whether positive or negative, to the classification result.
As we describe in \Cref{sec:results}, there is no unique pattern common across all the communities, even though some features are represented many times in these plots.

\section{Additional Robustness Checks}
\label{app:Robustness}
We employed TabPFN~\cite{hollmann2025accurate}, a state-of-the-art machine learning architecture, to further validate the robustness of our results.
To meet TabPFN's input constraints, we downsampled each dataset to approximately \num{10000} instances.
We then replicated the classification pipelines from the main experiments using this model.
The results closely matched those obtained with the Random Forest classifier, reinforcing the meaningfulness of the psycholinguistic embeddings.

\subsection{TabPFN Results}
\label{app:tabpfn}
In \Cref{fig:b5} we show the results of the experiments, disaggregated by the activity thresholds defined in \Cref{sec:Methods}. 
We find the results to be in line with the ones obtained with a Random Forest model in the experiments presented in \Cref{sec:results}.
Once again, we show how the volume of activity in \rconsp only marginally influences the accuracy of the classification.

\subsection{Computational Resources}
We ran our experiments on a machine with: a 64-core CPU, 256 GB of RAM, and an NVIDIA A100 GPU. 
The dataset filtering required a week of execution.
The feature extraction required five days of execution.
Model training and evaluation required a week of execution.
We leave the repository at the following link: https://anonymous.4open.science/r/reddit\_ct-EF52/

\subsection{Description of important features}
We report here the descriptions of the notable features we showed in \Cref{tab:summary_feat}, extracting them from the LIWC22 documentation~\cite{boyd2022development}.
The following descriptions contain examples from the categories, which can be swears or offensive words. Readers 
\begin{squishlist}
    \item \textbf{filler}: This category captures conversational filler words. Frequently used examples include words like ``I mean'', ``wow'', ``sooo'', and ``youknow''.
    \item \textbf{WC (Word Count)}: This is a summary variable that simply represents the total word count of the analyzed text.
    \item \textbf{sexual}: This is an expanded dictionary category that captures sexual language. The most frequently used exemplars include terms like ``sex'', ``gay'', ``pregnan*'', and ``dick''.
    \item \textbf{swear}: This category captures swear words, with frequent examples including ``shit'', ``fuckin'', ``fuck'', and ``damn''. In LIWC-22, swear words are conceptualized as part of the overall affect and tone dictionaries (rather than strictly negative emotion), as their usage has evolved and is now just as likely to reflect positive sentiment in informal contexts.
    \item \textbf{illness}: A subcategory of the broader Health dimension, this captures disease names and physical symptoms related to illness. Common examples include ``hospital'', ``cancer*'', ``sick'', and ``pain''.
    \item \textbf{death}: This category captures language referring to mortality. Frequently used words include ``death'', ``dead'', ``die'', and ``kill''.
    \item \textbf{emo\_anger}: This represents the emotion of anger, restricted strictly to true emotion labels or words that strongly imply the emotion. Examples include ``hate'', ``mad'', ``angry'', and ``frustr*''.
    \item \textbf{emo\_anx}: This represents the emotion of anxiety, similarly restricted to clear emotion labels and strong implications of anxiety. Frequently used exemplars include ``worry'', ``fear'', ``afraid'', and ``nervous''.
    \item \textbf{mental}: This is a mental health category that typically identifies psychological diagnoses (e.g., ``bipolar'', ``neurosis'') or related behaviors (e.g., ``suicide'', ``addiction''). Other frequent examples include ``depressed'' and ``trauma''.
    \item \textbf{curiosity}: Falling under the ``Motives'' dimension, these words reflect an author's search for or interest in new knowledge or experiences, which is thought to correlate with the personality trait of openness. Examples include ``scien*'', ``look for'', ``research'', and ``wonder''.
    \item \textbf{achieve}: Part of the psychological ``Drives'' dimension, this category captures language related to achievement. Frequently used examples include ``work'', ``better'', ``best'', and ``working''.
    \item \textbf{clout}: This is one of the four summary variables in LIWC-22, representing the language of leadership and social status.
\end{squishlist}

\clearpage

\begin{figure*}
    \centering
    \includegraphics[width=0.9\linewidth]{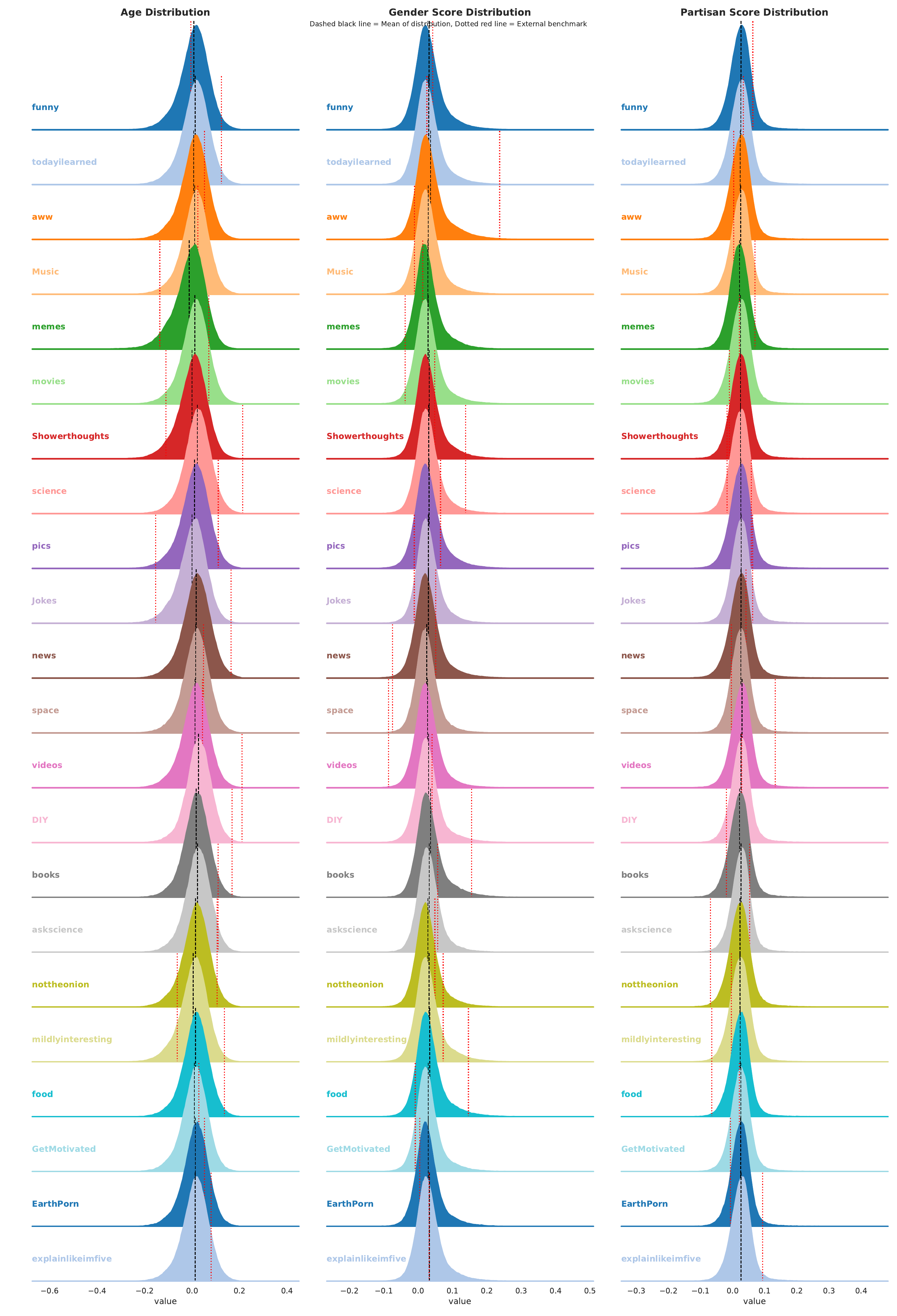}
    \caption{Distribution the socio-demographic scores of conspiracy users active in each mainstream subreddit. Black Dashed Line: Conspiracy mean, Red Dashed Line: Subreddit Score Mean, provided by~\cite{waller2021quantifying}}
    \label{fig:b_10}
\end{figure*}

\begin{table*}[ht]
\caption{Data sizes for each subreddit, in millions of comments. We consider the comments of users who have made at least 20 contributions to the subreddit.}
\label{tab:sizes_comments}
\centering
\begin{tabular}{lSSS}
\toprule
\text{Subreddit} & \text{\# comments of CT users} & \text{\# comments of NOCT users} & \text{Total} \\
\midrule
r/memes& 5.67 & 76.65 & 82.32 \\
r/todayilearned & 17.58 & 39.11 & 46.69 \\
r/Showerthoughts & 6.04 & 23.38 & 29.42 \\
r/nottheonion & 3.99 & 8.74 & 12.73 \\
r/aww & 5.29 & 29.0 & 34.29 \\
r/Music & 3.23 & 12.01 & 15.24 \\
r/movies & 12.41 & 37.76 & 50.17 \\
r/science & 3.67 & 7.94 & 10.61 \\
r/pics & 24.31 & 68.76 & 93.07 \\
r/Jokes& 1.71 & 7.55 & 9.26 \\
r/news & 24.54 & 41.38 & 65.96 \\
r/videos& 14.82 & 32.93 & 47.75 \\
r/space & 1.77 & 4.06 & 5.83 \\
r/askscience & 0.80 & 2.8 & 3.6 \\
r/DIY & 0.88 & 3.51 & 4.39 \\
r/books & 1.71 & 8.35 & 10.06 \\
r/mildlyinteresting & 5.26 & 21.2 & 26.46 \\
r/food & 1.37 & 6.48 & 7.85 \\
r/EarthPorn & 0.6 & 2.76  & 3.36 \\
r/GetMotivated & 0.65 & 2.18 & 2.83 \\
r/explainlikeimfive & 3.52 & 11.23 & 14.75 \\
r/funny & 22.56 & 76.26 & 88.82 \\
\bottomrule
\end{tabular}
\end{table*}

\begin{table*}
\caption{Dataset sizes for each mainstream subreddit in number of users. The datasets are balanced, so the number of conspiracy users is (approximately) equal to the number of non-conspiracy users.}
\sisetup{table-number-alignment=center,table-format=6.0}
\centering
\begin{tabular}{lSSSSS}
\toprule
Subreddit & \text{(0,1]} & \text{(1,10)} & \text{[10,100)} & \text{[100,$\infty$)} & \text{Total} \\
\midrule
r/funny               & 88362  & 121808 & 74412  & 23250  & 307832 \\
r/memes               & 24195  & 35208  & 23456  & 6072   & 88931  \\
r/science             & 15564  & 26124  & 19187  & 8443   & 69318  \\
r/Music               & 15849  & 24982  & 18147  & 6674   & 65652  \\
r/todayilearned       & 65644  & 99640  & 64532  & 21908  & 251724 \\
r/aww                 & 29886  & 43720  & 27650  & 8719   & 109975 \\
r/movies              & 41689  & 62096  & 41176  & 14283  & 159244 \\
r/Showerthoughts      & 29506  & 46913  & 32209  & 10140  & 118768 \\
r/pics                & 94883  & 137573 & 85882  & 26635  & 344973 \\
r/Jokes               & 8322   & 12326  & 8688   & 3114   & 32450  \\
r/news                & 69170  & 110814 & 75465  & 27734  & 283183 \\
r/videos              & 53849  & 80752  & 52473  & 17093  & 204167 \\
r/space               & 6856   & 11245  & 8394   & 3416   & 29911  \\
r/askscience          & 3011   & 4662   & 3056   & 1182   & 11911  \\
r/DIY                 & 3860   & 5796   & 3872   & 1544   & 15072  \\
r/books               & 8326   & 12141  & 8507   & 3061   & 32035  \\
r/nottheonion         & 18078  & 31548  & 23808  & 8832   & 82266  \\
r/mildlyinteresting   & 28254  & 43890  & 30176  & 9734   & 112054 \\
r/food                & 6649   & 10356  & 7048   & 2598   & 26651  \\
r/EarthPorn           & 2158   & 3412   & 2302   & 949     & 8821   \\
r/GetMotivated        & 2321   & 3788   & 3059   & 1154   & 10322  \\
r/explainlikeimfive   & 14155  & 21875  & 15700  & 5680   & 57410  \\
\bottomrule

\end{tabular}
\end{table*}

% \label{app:features}

\end{document}